\documentclass{article}
\usepackage[utf8]{inputenc}
\usepackage{authblk}
\usepackage{caption}
\usepackage{subcaption}
\usepackage{graphicx}
\usepackage{listings}
\usepackage{amssymb}
\usepackage{amsmath}
\usepackage{booktabs} 
\usepackage[dvipsnames]{xcolor}
\usepackage{multirow}
\usepackage[a4paper, total={6.5in, 9.5in}]{geometry}

\newcommand{\beq}{\begin{equation}}
\newcommand{\eeq}{\end{equation}}
\newcommand{\overliner}{\begin{array}}
\newcommand{\earr}{\end{array}}
\newcommand{\beqarr}{\begin{eqnarray}}
\newcommand{\eeqarr}{\end{eqnarray}}
\newcommand{\beqar}{\begin{eqnarray*}}
\newcommand{\eeqar}{\end{eqnarray*}}
\newcommand{\bef}{\begin{figure}}
\newcommand{\eef}{\end{figure}}

\newcommand{\bm}[1]{\mbox{\boldmath$#1$}}

\title{Multirotor-assisted measurements of wind-induced drift of irregularly shaped objects in aquatic environments}
\date{}

\author[1]{Javier Gonz\'alez-Rocha\footnote{Corresponding author at: Department of Mechanical Engineering, University of California Riverside,
900 University Ave.
Riverside, CA 92521, United States. \newline \indent \textit{E-mail address:} javier.gonzalezrocha@ucr.edu}}
\author[2]{Alejandro J. Sosa}
\author[3]{Regina Hanlon}
\author[4]{Arthur A. Allen}
\author[5]{Irina Rypina}
\author[3]{David G. Schmale III}
\author[2]{Shane D. Ross}

\affil[1]{\small Department of Mechanical Engineering, University of California, Riverside, CA, United States}
\affil[2]{Kevin T. Crofton Department of Aerospace and Ocean Engineering, Virginia Tech, VA, United States}
\affil[3]{School of Plant and Environmental Sciences, Virginia Tech, VA, United States}
\affil[4]{US Coast Guard, Office of Search and Rescue, CT, United States}
\affil[5]{Physical Oceanography Department, Woods Hole Oceanographic Institution, Woods Hole, MA, United States}

\begin{document}

\maketitle

\begin{abstract}
Ocean hazardous spills and search and rescue incidents are becoming more prevalent as maritime activities increase across all sectors of society. However, emergency response time remains a factor due to the lack of information available to accurately forecast the location of small objects. Existing drifting characterization techniques are limited to objects whose drifting properties are not affected by on-board wind and surface current sensors. To address this challenge, we study the application of multirotor small unmanned aerial systems (sUAS), and embedded navigation technology, for on-demand wind velocity and surface flow measurements to characterize drifting properties of small objects. An off-the-shelf quadrotor was used to measure the speed and direction of the wind field at 10 m above surface level near a drifting object. We also leveraged sUAS-grade attitude and heading reference systems and GPS antennas to build water-proof tracking modules that record the position and orientation, as well of translational and rotational velocities, of objects drifting in water. The quadrotor and water-proof tracking modules were deployed during field experiments conducted in Claytor Lake, VA and the Atlantic Ocean south of Martha's Vineyard, MA to characterize the leeway parameters of manikins simulating a person in water. Leeway parameters were found to be an order of magnitude within previous estimates that were derived using conventional wind and surface current observations. We also determined that multirotor sUAS and water-proof tracking modules can provide accurate and high-resolution ambient information that is critical to understand how changes in orientation affect the downwind displacement and jibing characteristics of small objects floating in water. These findings support further development and application of multirotor sUAS technology for leeway characterization and understanding the effect of an object's downwind-relative orientation on its drifting characteristics.
\end{abstract}


\section{Introduction}
\label{s:introduction}
Hazardous spills and search-and-rescue events have increased with more human activity present in ocean environments~\cite{sturmer2020japanese,cook2020intensive,money20201marine}. Efforts to improve emergency response rely on trajectory models with low uncertainty bounds to localize drifting objects in minimum time~\cite{breivik2011wind,wang2015drifting}. However, the uncertainty associated with search regions derived from trajectory models is limited by the accuracy of drift properties available for irregularly-shaped objects (e.g., persons in distress, hazardous material, accident debris, plastic waste, and disabled vessels)~\cite{brushett2014determining,Brushett2017application}. Moreover, drift information only exist for a relative small number of objects due to the time and cost involved to estimate drift parameters using conventional position tracking, wind, and surface flow sensors. Therefore, new techniques to characterize drift properties employing multirotor small unmanned aircraft systems (sUAS) technology can help improve the uncertainty bounds of trajectory models, particularly for those test objects unable to house a suite of sensors.

How floating objects move in the ocean under varied atmospheric and surface current conditions is characterized
by leeway parameters. As described in \cite{breivik2011wind}, the leeway of an object is the motion resulting from wind and wave forcing relative to the ocean's surface current. In practice, leeway parameters are estimated using inertial position, atmospheric wind and surface current measurements collected from an assortment of sensors during field experiments~\cite{breivik2011wind,allen2005leeway}. Typically, position tracking of drifting objects is conducted with on-board GPS antennas. The speed and direction of atmospheric wind is extrapolated to a height of 10 meters above sea level (ASL) using observations from in-situ sensors mounted on a drifting object or nearby remote sensors, and in special instances, leveraging high-resolution  weather models. The surface current near an object is measured directly using a flowmeter or indirectly using a GPS-equipped surface drifter, and in unique situations, employing remote sensors as well. However, for the common scenario, where only in-situ observations of wind and surface flow are feasible, reliable leeway experiments are only possible for objects whose drift properties are unaltered by on-board sensors. Hence, exploring new alternatives to measure wind and surface current is necessary to expand the leeway information available for small and irregularly-shaped objects whose footprint is smaller than 3 meters.  

Employing multirotor sUAS technology to measure wind and surface currents can help overcome limitations of leeway characterization methods. Already, multirotor sUAS have been leveraged for measuring atmospheric parameters in studies related to atmospheric transport~\cite{nolan2018understanding,nolan2018coordinated}, weather forecasting~\cite{Varentsov2019experience,greene2019environmental,barbieri2019small}, and climate change~\cite{bates2013measurements,araus2018breeding,varela2019assessing}. These systems are low cost, portable, mobile, and easy to deploy and recover from constrained environments, which is ideal for on-demand and targeted wind observations near a drifting object~\cite{gonzalez2019sensing}. Moreover, the navigation system on board a multirotor sUAS, which includes an attitude and heading reference system (AHRS) and GPS antenna, can be re-purposed for position and orientation tracking of drifting objects. Combined, these adaptations of UAS technology during field experiments can facilitate the expansion of leeway data bases used as inputs for trajectory forecasting models predicting low-uncertainty search regions for a broad class of objects.

In this paper we present a new method for measuring the ambient parameters required to characterize leeway properties of small objects using multirotor UAS technology. The method involves using a quadrotor for conducting wind field observation near drifting objects at 10 meters above sea level (ASL). Additionally, tracking modules that were built with multirotor UAS navigation technology are used to record high-resolution measurements of position, attitude and heading, as well as translational and rotational velocities, for drifting objects. The effectiveness of these measurement techniques was tested in field experiments conducted both in lake and ocean environments. Results from leeway experiments demonstrate the potential of the presented measurement method to characterize leeway properties for small and irregularly-shaped objects. 

The organization of this paper is as follows. In Section~\ref{s:methodology} the analysis, experiments, and hardware used to estimate leeway parameters are described. Section~\ref{s:Leeway_Experiments_Results} presents a comprehensive summary of leeway experiments that includes wind velocity, object tracking, and leeway parameter estimation results. A discussion of leeway experiment results and the potential of observations derived from multirotor UAS technology to streamline leeway parameter estimation is presented in Section~\ref{s:Discussion}. Finally, conclusions and direction of future work are presented in Section~\ref{s:conclusion}.

\section{Methodology}
\label{s:methodology}
\subsection{Leeway Analysis}
\label{ss:windage_analysis}
\begin{figure}
 \centering
         \includegraphics[width=7 cm]{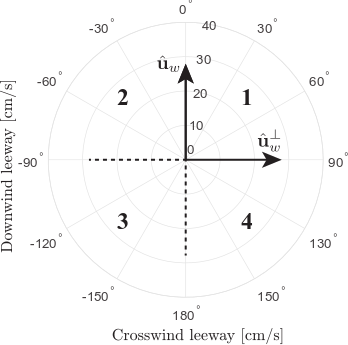}
    \caption{\footnotesize The four quadrants of the two-dimensional leeway space. The axis are the downwind and crosswind unit vectors, $\hat{\mathbf{u}}_w^{\perp}$ and $\hat{\mathbf{u}}_w$, respectively. For example, quadrant 2 has positive downwind leeway and negative crosswind leeway.
    }\label{fig:leeway_polar_diagram}
         \label{subfig:Claytor_Lake_Experiments}
\end{figure}

A basic formulation of the motion of an object floating at the air-water interface is to consider the change in the object's two-dimensional (horizontal) position with time, $\mathbf{x}(t)$, via the ordinary differential equation,
\begin{equation}
    \frac{d}{dt}\mathbf{x} = \mathbf{u}(\mathbf{x},t) 
\end{equation}
where $\mathbf{u}(\mathbf{x},t)$ is the two-dimensional hybrid surface velocity 
\cite{allshouse2017impact}. 
Several models for representing the hybrid surface velocity exist.  For instance, based on \cite{breivik2011wind} and \cite{beron2016inertia}, we consider a model of the form,
\begin{equation}\label{general_model}
\begin{split}
    \mathbf{u}(\mathbf{x},t) =  \mathbf{u}_c(\mathbf{x},t)
    + C_w\mathbf{u}_w(\mathbf{x},t) +  C_w^{\perp}\mathbf{u}_w^{\perp}(\mathbf{x},t)+\cdots\\
     \mathbf{b}_w(\hat{\mathbf{u}}_w(\mathbf{x},t),\hat{\mathbf{u}}_w^{\perp}(\mathbf{x},t)) + \bm{\varepsilon}_w(\hat{\mathbf{u}}_w(\mathbf{x},t),\hat{\mathbf{u}}_w^{\perp}(\mathbf{x},t))
    \end{split}
\end{equation}
where $\mathbf{u}_c(\mathbf{x},t)$ is the ambient water current velocity (i.e., the average velocity between 0.3 and 1.0 m depth), $\mathbf{u}_w(\mathbf{x},t)$ is the wind velocity (at 10 m height above the water surface), $C_w$ is the leeway coefficient in the downwind direction (positive downwind), $C_w^{\perp}$ is the leeway coefficient in the cross-wind direction (positive to the right), 
and the offset vector can be written in terms of downwind and cross-wind components,
\begin{equation}
\mathbf{b}_w((\hat{\mathbf{u}}_w(\mathbf{x},t)),\hat{\mathbf{u}}_w^{\perp}(\mathbf{x},t))=
\mathrm{b}_w        \hat{\mathbf{u}}_w        (\mathbf{x},t) +
\mathrm{b}_w^{\perp}\hat{\mathbf{u}}_w^{\perp}(\mathbf{x},t),
\end{equation}
and the (stochastic) error vector similarly can be written,
\begin{equation}
\bm{\varepsilon}_w(\hat{\mathbf{u}}_w(\mathbf{x},t),\hat{\mathbf{u}}_w^\perp(\mathbf{x},t))=
\varepsilon_w          \hat{\mathbf{u}}_w        (\mathbf{x},t) + 
\varepsilon_w^{\perp}  \hat{\mathbf{u}}_w^{\perp}(\mathbf{x},t),
\end{equation}
where the (normalized unit) downwind direction is $\hat{\mathbf{u}}_w(\mathbf{x},t)$, i.e., $\mathbf{u}_w(\mathbf{x},t) = u_w(\mathbf{x},t) \hat{\mathbf{u}}_w(\mathbf{x},t)$, where $u_w(\mathbf{x},t)= \| \mathbf{u}_w(\mathbf{x},t) \|$.
Note that the cross-wind vector, $\mathbf{u}_w^{\perp}(\mathbf{x},t)=\mathbf{\Omega}\mathbf{u}_w (\mathbf{x},t)$,
is the wind velocity rotated 90$^{\circ}$ to the right of the downwind direction, where
$\mathbf{\Omega}$ is the matrix providing a 90$^{\circ}$ clockwise rotation,
\begin{equation}
\mathbf{\Omega} =
\begin{bmatrix}
~~ 0 & 1 \\
-1 & 0
\end{bmatrix}.
\end{equation}
Note that $C_w$, $C_w^{\perp}$, $b_w$, $b_w^{\perp}$, $\varepsilon_w$, and $\varepsilon_w^{\perp}$ are all scalars.  For this situation, the velocity of the object relative to the ambient water current is the leeway velocity, $\mathbf{u}_{l}(\mathbf{x},t) \equiv \mathbf{u}(\mathbf{x},t)-\mathbf{u}_c(\mathbf{x},t)$, is given by, \begin{equation}
  \mathbf{u}_l(\mathbf{x},t)=C_w\mathbf{u}_w (\mathbf{x},t)+C_w^{\perp}\mathbf{u}_w^{\perp} (\mathbf{x},t)+ \mathbf{b}_w(\hat{\mathbf{u}}_w(\mathbf{x},t),\hat{\mathbf{u}}_w^\perp(\mathbf{x},t)) + \bm{\varepsilon}_w(\hat{\mathbf{u}}_w(\mathbf{x},t),\hat{\mathbf{u}}_w^\perp(\mathbf{x},t)),
  \label{simple_model}
\end{equation}
based on \eqref{general_model}.
If we can measure the relative velocity, $\mathbf{u}_l(\mathbf{x},t)$, and the 10 m height wind velocity, $\mathbf{u}_w(\mathbf{x},t)$, as in \cite{breivik2011wind}, then we can estimate the coefficients $C_w$ and $C_w^{\perp}$ via linear regression,
\begin{equation}
\begin{split}
\rm{Downwind~leeway}: \quad
\mathbf{u}_l(\mathbf{x},t)\cdot\hat{\mathbf{u}}_w(\mathbf{x},t) 
&= C_w u_w + \mathrm{b}_w + \varepsilon_w,  \\
\rm{Crosswind~leeway}: \quad \mathbf{u}_l(\mathbf{x},t)\cdot\hat{\mathbf{u}}_w^\perp(\mathbf{x},t) 
&= 
C_w^{\perp}  u_w  + \mathrm{b}_w^{\perp} + \varepsilon_w^{\perp}.
\end{split}
\label{DWL_CWL}
\end{equation}
By measuring the leeway motion of an object and comparing with the wind speed ($u_w$),
constrained and unconstrained linear regression provides the slopes 
($C_w$ and $C_w^{\perp}$), the offsets 
($b_w$ and $b_w^{\perp}$), 
and a measure of uncertainty/error ($\varepsilon_w$ and $\varepsilon_w^{\perp}$), usually the standard deviation \cite{breivik2011wind}. The difference between constrained and unconstrained approaches is that the former produces a linear fit passing through the origin such that the offset is equal to zero. Moreover, in previous studies \cite{allen2005leeway, breivik2011wind}, the drift coefficients for crosswind have been decomposed into those for positive $(+)$  and negative $(-)$ crosswind behaviors, that is, allowing for non-symmetrical drift coefficients, left and right-drifting objects moving differently. In this study, we further allow for the possibility of both positive $(+)$ and negative $(-)$ downwind behaviors. Our expanded set of leeway equations is then as follows, where the $\mathbf{x}$ and $t$ dependence of $\mathbf{u}_l$, $\hat{\mathbf{u}}_w$, and $\hat{\mathbf{u}}_w^\perp$ is understood,
\begin{equation}
\begin{split}
\rm{Positive~downwind~leeway}: \quad \rm{if}~\mathbf{u}_l\cdot\hat{\mathbf{u}}_w > 0, \quad 
\mathbf{u}_l\cdot\hat{\mathbf{u}}_w
&= C_w^+ u_w + \mathrm{b}_w^+ + \varepsilon_w^+,   \\
\rm{Negative~downwind~leeway}: \quad \rm{if}~\mathbf{u}_l\cdot\hat{\mathbf{u}}_w < 0, \quad 
\mathbf{u}_l\cdot\hat{\mathbf{u}}_w
&= C_w^- u_w + \mathrm{b}_w^- + \varepsilon_w^-,   \\
\rm{Positive~crosswind~leeway}: \quad \rm{if}~\mathbf{u}_l\cdot\hat{\mathbf{u}}_w^{\perp} > 0, \quad 
\mathbf{u}_l\cdot\hat{\mathbf{u}}_w^\perp 
&= C_w^{\perp+}  u_w  + \mathrm{b}_w^{\perp+} + \varepsilon_w^{\perp+}, \\
\rm{Negative~crosswind~leeway}: \quad \rm{if}~\mathbf{u}_l\cdot\hat{\mathbf{u}}_w^{\perp} < 0, \quad 
\mathbf{u}_l\cdot\hat{\mathbf{u}}_w^\perp 
&= C_w^{\perp-}  u_w  + \mathrm{b}_w^{\perp-} + \varepsilon_w^{\perp-},
\end{split}
\label{DWL_CWL_pos_neg}
\end{equation}
Given the different combinations of positive and negative downwind and crosswind leeway, we will summarize results for each of the four quadrants of the leeway space, as shown schematically in Figure \ref{fig:leeway_polar_diagram}.

We note that in Allshouse et al.~\cite{allshouse2017impact}, they simplified the leeway analysis by setting $C_w^{\perp}$, $\mathbf{b}_w$, and $\mathbf{\varepsilon}_w$ as being negligible, which gives the simplest hybrid model. Given the uncertainties in forecast models of $\mathbf{u}_c(\mathbf{x},t)$ and $\mathbf{u}_w(\mathbf{x},t)$, as well as the variability in $C_w$ even for the same class of object \cite{breivik2011wind}, the simplest model may be  appropriate in some circumstances. However, the standard approach for estimating and using leeway, as used by the U.S. Coast Guard \cite{allen2005leeway}, considers the more realistic model, \eqref{DWL_CWL_pos_neg}.

\subsection{Leeway Experiments}
\label{ss:leeway_experiments}

In this study, the leeway parameters $C_w$ and $C_w^{\perp}$ for a person-in-water scenario are calculated from experiments conducted both at Claytor Lake in Virginia and in coastal waters of the Atlantic ocean south-southeast of Martha's Vineyard, Massachusetts. The two sets of experiments consisted of releasing into the water, as shown in Figure~\ref{fig:leeway_experiment_setup}, pairs of surface drifters and search-and-rescue training manikins with integrated GPS devices for position tracking. The GPS positions of the surface drifters and the manikins were used to calculate the flow-relative velocity of each manikin, assuming that variation in water surface current is insignificant over the distance of the separation between the two. Additionally, a quadrotor UAS was employed to measure the velocity of the 10 m AGL wind which was influencing the manikin motion. The quadrotor UAS used a model-based wind sensing approach presented in \cite{gonzalez2019sensing}. Together, flow relative and wind velocity measurements were used to calculate the leeway parameters for a person-in-water scenario using the analysis presented in Section~\ref{ss:windage_analysis} for both the lake and ocean environments.  

\begin{figure}[h!]
     \centering
     \begin{subfigure}[b]{0.47\textwidth}
         \centering
         \includegraphics[width=\textwidth]{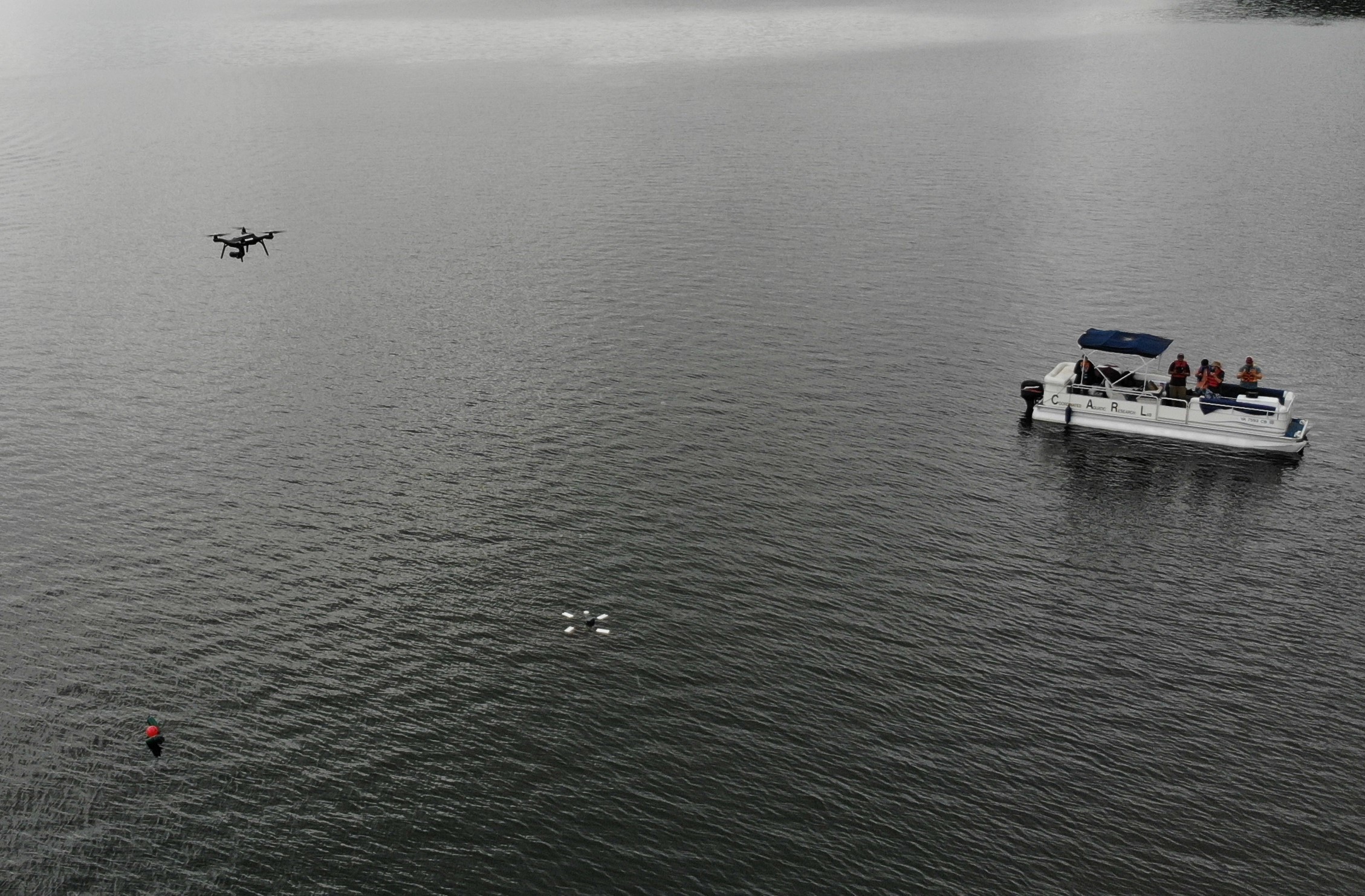}
         \caption{}
         \label{subfig:Claytor_Lake_Experiments}
     \end{subfigure} \hfill
     \begin{subfigure}[b]{0.47\textwidth}
         \centering
         \includegraphics[width=\textwidth]{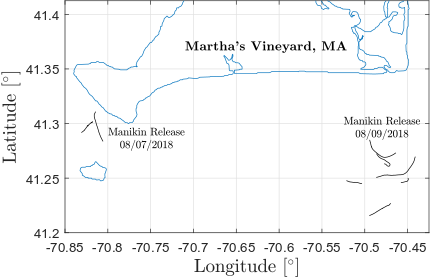}
         \caption{}
         \label{subfig:Ocean_Experiments}
     \end{subfigure}
        \caption{a) An image of leeway experiments conducted in Claytor Lake, VA on June 25th, 2018. b)  A map of leeway experiments conducter in the ocean south of Martha's Vineyard, MA on August 7th and  9th, 2018.}
        \label{fig:Leeway_Experiments}
\end{figure}

\begin{figure}[h!]
 \centering
         \includegraphics[width=13 cm]{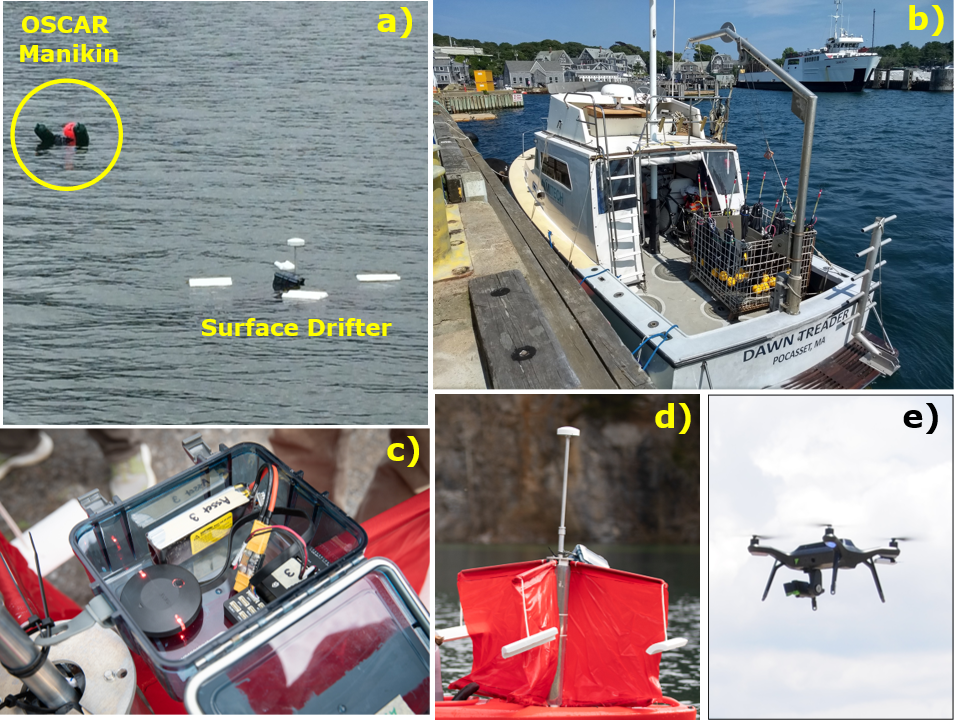}
     \caption{ a) Field experiments conducted to determine the leeway of a person in water using ambient current and wind velocity measurements obtained from a surface drifter and multirotor sUAS.  b) Maritime research vessel used to base ocean leeway experiments. c) Pixhawk-GPS module configured to track navigation information at 5 Hz. d) CODE/DAVIS drifter used to measure ambient flow. e) 3DR Solo quadrotor used to measure wind velocity 10 m wind measurements.  
    }\label{fig:leeway_experiment_setup}
\end{figure}

\subsection{Search and Rescue Training Manikin}
\label{ss:search_and_rescue_training_manikin}

\begin{figure}[h!]
    \centering
     \begin{subfigure}[b]{0.47\textwidth}
         \centering
         \includegraphics[width=\textwidth]{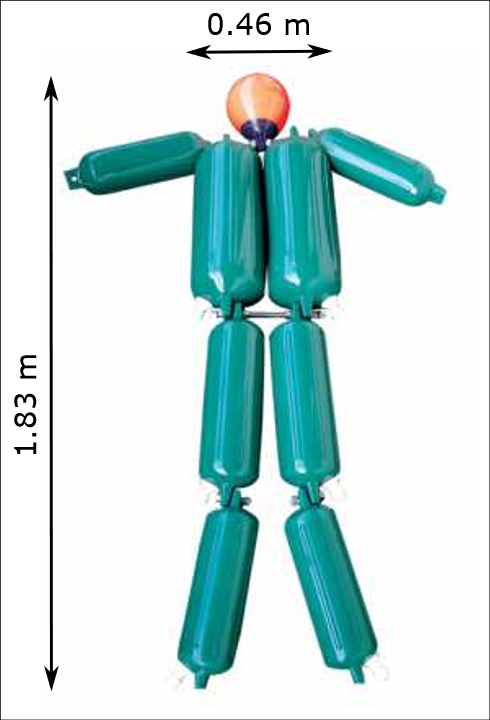}
         \caption{}
         \label{subfig:OSCAR_manikin}
     \end{subfigure}
         \begin{subfigure}[b]{0.47\textwidth}
         \centering
         \includegraphics[width=\textwidth]{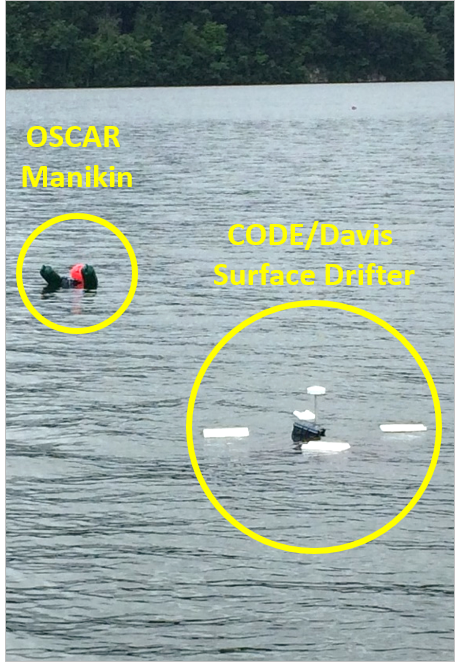}
         \caption{}
         \label{subfig:OSCAR_and_surface_drifter}
     \end{subfigure}
    \caption{ a) A schematic drawing showing the dimensions of the OSCAR search-and-rescue manikin used during leeway experiments as a PIW surrogate.  b) An image of the OSCAR search-and-rescue manikin drifting next to a CODE/DAVIS surface drifter. 
    }\label{fig:OSCAR}
\end{figure}

The manikins used during lake and ocean leeway experiments are the OSCAR dummy manufactured by Emerald Marine Products for person-in-water rescue training. As shown in Figure \ref{fig:OSCAR}, the manikins are each configured from eight heavy-duty vinyl bladders with fill/drain fittings, six stainless steel joints, and two galvanized lifting shackles. Fully extended and filled with water, the manikins weigh 82~kg, are 2~m tall, and have a chest width of 0.5~m. Prior to leeway experiments, each manikin was filled with sufficient water to achieve an upright floating position at chest level. In this configuration, the back and front as well as the left and right sides of each manikin are symmetrical. 

\subsection{Ambient Sensing Hardware}
\label{ss:ambient_sensing_hardware}

\subsubsection{Surface Drifter}
\begin{figure}[h!]
    \centering
     \begin{subfigure}[b]{0.47\textwidth}
         \centering
         \includegraphics[width=\textwidth]{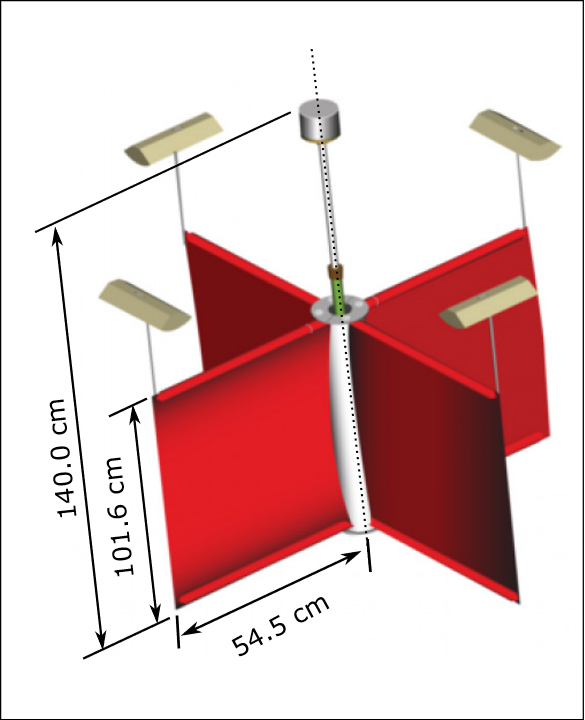}
         \caption{}
         \label{subfig:CODEdrifter_schematic}
     \end{subfigure}
         \begin{subfigure}[b]{0.47\textwidth}
         \centering
         \includegraphics[width=\textwidth]{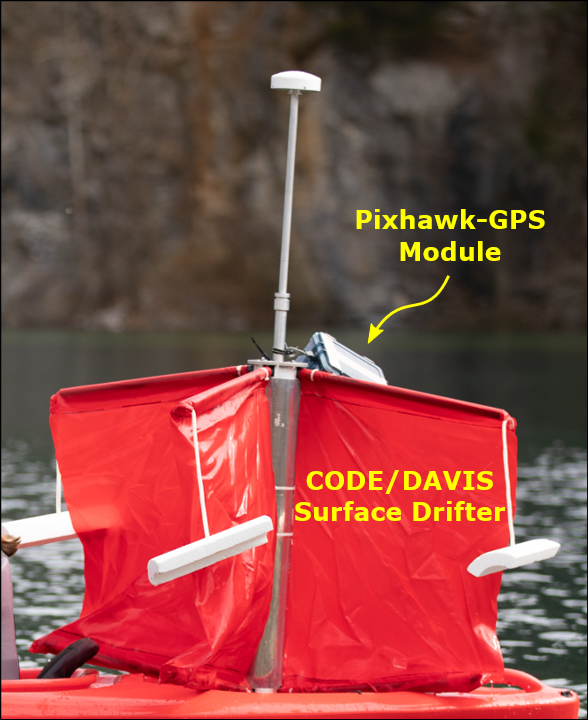}
         \caption{}
         \label{subfig:CODEdrifter_schematic_labelr}
     \end{subfigure}
\caption{ a) A schematic with dimensions of the CODE/DAVIS surface drifter used to measure the water current in a 1-meter column below the surface. b) A CODE/Davis surface and Pixhawk-GPS module integration. 
    }\label{fig:surface_drifter}
\end{figure}

The surface current surrounding the search and rescue training manikin during field experiments was measured by tracking the position of a surface drifter with a water-proof module as show in Figure~\ref{fig:surface_drifter}. Fully deployed, surface drifters have a height of 1 m and a width of 1.4 m. This type of drifter is certified to CODE (Coastal Ocean Dynamic Experiment Standards) to drift with estuary and ocean surface currents extending a meter below the surface level. The drifter's position was tracked relative to a north-east-down reference frame using GPS antennas tied to the center shaft of the surface drifter as shown in Figure~\ref{fig:surface_drifter}. Position information from GPS was used to determine velocity of surface currents surrounding the search and rescue training manikin.

\subsubsection{GPS Position Tracking }
\label{ss:GPS_position_tracking}

Two types of GPS receivers were used to track the position of manikin-and-drifter pairs during lake and ocean leeway experiments. For leeway experiments conducted in Claytor Lake, built-in-house GPS modules were used to track search-and-rescue manikins and surface drifters. These GPS modules, referred to as the Pixhawk-GPS modules, were developed using sUAS-grade attitude and heading reference systems and GPS antennas. Experiments conducted at sea incorporated both the Pixhawk-GPS modules and off-the-shelf SPOT Trace GPS units to track search-and-rescue manikins. Surface drifters, on the other hand, were solely integrated with SPOT Trace GPS units. The sensor configuration employed during ocean experiments provided redundancy and allowed for the performance of the two GPS modules to be compared. 

\subsubsection{Pixhawk-GPS Modules}
\label{ss:pixhawk-GPS_modules}
Search-and-rescue manikins were integrated with waterproof Pixhawk-GPS modules that were developed in house to track the position of objects drifting in aquatic environments. Each Pixhawk-GPS module has a 3s 11.1 Volts Lithium-Polymer (LiPo) battery that can last up to 6 hours, a Pixhawk autopilot computer and a 3DR SiK 900Hz radio for short-range telemetry, and a HERE GPS antenna for position tracking. While operating, each Pixhawk-GPS module records position information with an update rate of 5 Hz. This resolution in position information may potentially provide new information for understanding small-scale dynamic effects experienced by drifting assets in aquatic environments. However, for the leeway analysis presented in this paper, position information from Pixhawk-GPS modules was processed using a 5-minute moving average. 

The magnetic heading as well as the translational and rotational motions of each manikin were also measured leveraging the attitude and heading reference system inside of each Pixhawk flight autopilot. Each attitude and heading reference system has three sets of compasses, accelerometers, and gyroscopes that sample measurements with the update rates shown in Table~\ref{table:pixhawk-integrated_sensors}. On board each Pixhawk, attitude and heading reference system measurements are processed in real time using an extended Kalman filter to correct for sensor noise and biases. These additional features of Pixhawk-GPS modules were exploited to measure the downwind-relative heading of each manikin as well as the wave period.


\begin{table}[tbh!]
\caption{ Summary of measurements recorded on-board Pixhawk-GPS modules.}
\centering \small
  \begin{tabular}{ccccccccccccccc}
    \toprule
    \multirow{2}{*}{ Measurement} &
      \multicolumn{2}{c}{ Direct}&& \multicolumn{2}{c}{ Estimated}\\
      \cline{2-3} \cline{5-6}  & Sensor & Rate  && Estimator & Rate    \\
      \midrule \midrule
Horizontal Position & GPS & 5 Hz & &EKF&  20 Hz   \\ 
Heading &Magnetometer& 5 Hz && EKF & 20 Hz \\
Transnational Rates & GPS & 5 Hz && EKF & 20 Hz     \\
 Angular Rates   & Gyroscope & 50 Hz && --- & ---    \\ \midrule
  \bottomrule
  \end{tabular}

  \label{table:pixhawk-integrated_sensors}
  \end{table}

\begin{figure}[h!]
    \centering
     \begin{subfigure}[b]{0.47\textwidth}
         \centering
         \includegraphics[width=\textwidth]{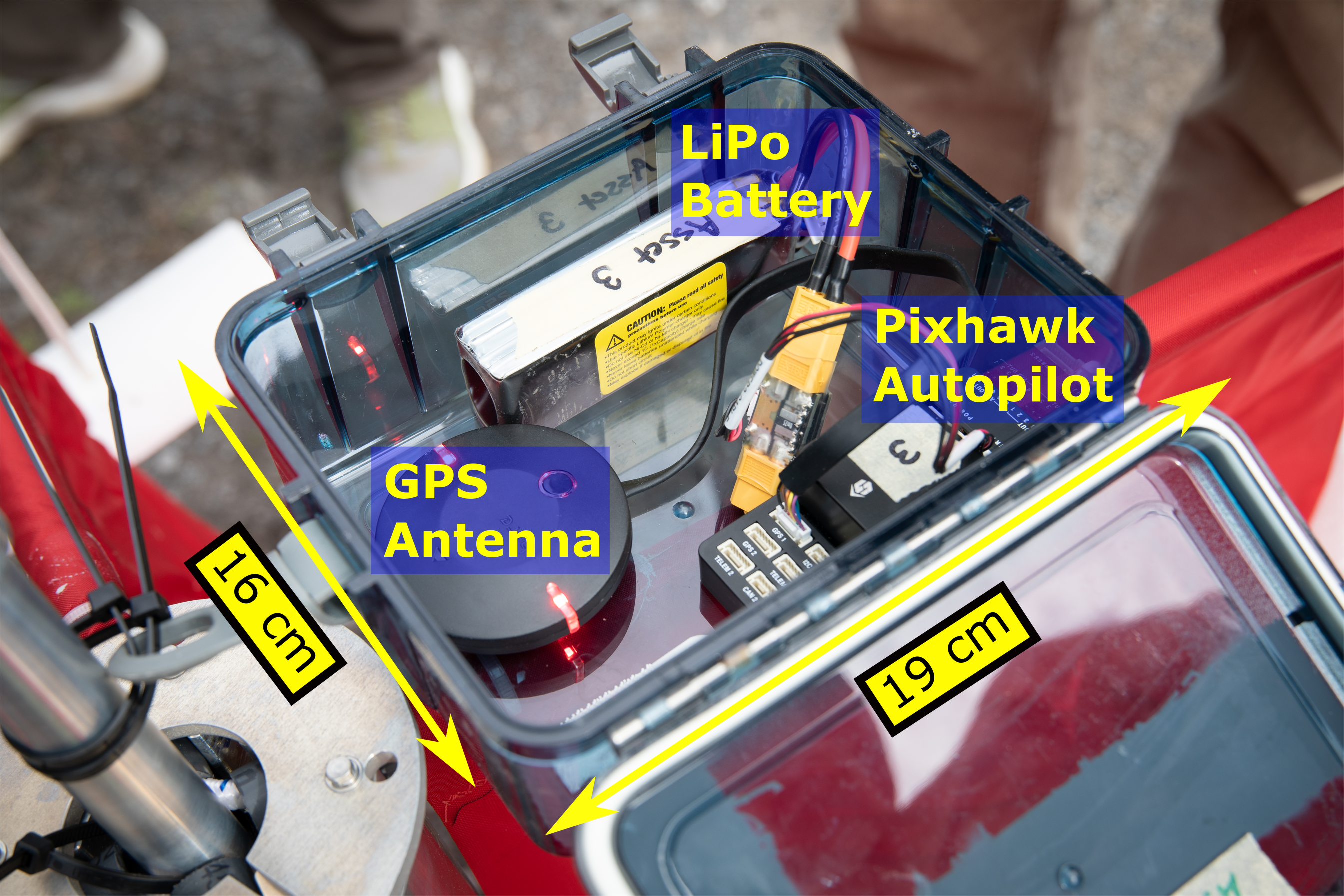}
         \caption{}
         \label{subfig:Pixhawk-GPS_module}
     \end{subfigure}
         \begin{subfigure}[b]{0.47\textwidth}
         \centering
         \includegraphics[width=\textwidth]{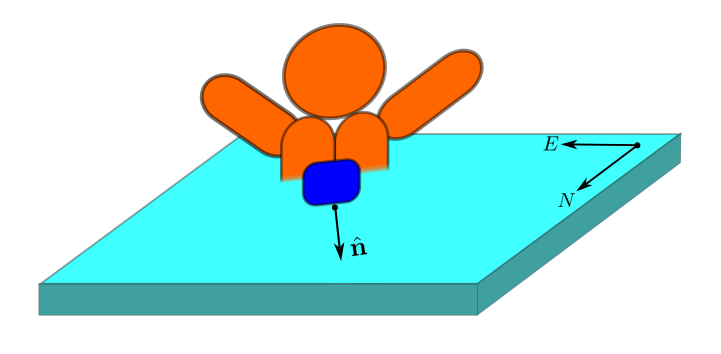}
         \caption{}
         \label{subfig:Manikin_Reference_Frame}
     \end{subfigure}
    \caption{ a) An image of the Pixhawk-GPS module used to track search-and-rescue manikins and accompanying surface drifters. b) A schematic drawing showing the front of search-and-rescue manikins and Pixhawk-GPS modules aligned such that the orientation of each manikin be measured as the angle between the unit vector $\hat{\mathbf{n}}$ and North. 
    }\label{fig:Pixhaw-GPS}
\end{figure}

\subsubsection{SPOT Trace GPS units}
\label{sss:SPOT_trace_GPS_units}

\begin{figure}[h!]
    \centering
     \begin{subfigure}[b]{0.47\textwidth}
         \centering
         \includegraphics[width=\textwidth]{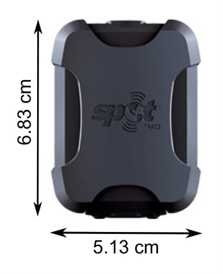}
         \caption{}
         \label{subfig:SPOTTraceGPS}
     \end{subfigure}
         \begin{subfigure}[b]{0.47\textwidth}
         \centering
         \includegraphics[width=\textwidth]{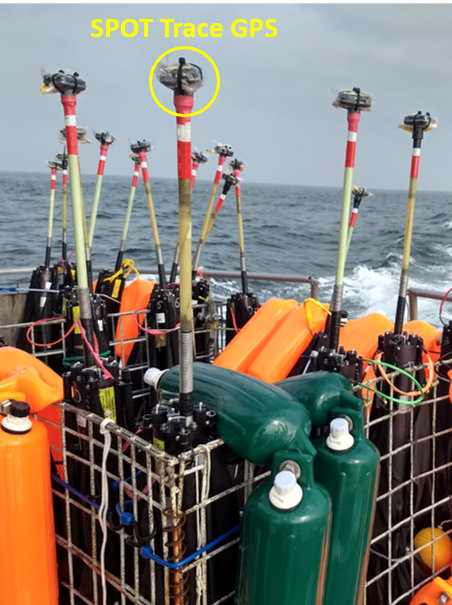}
         \caption{}
         \label{subfig:SPOTTrace_and_drifter}
     \end{subfigure}
   \caption{  a) A schematic with dimensions of the SPOT Trace GPS unit used to track the position of search-and-rescue manikins and accompanying surface drifters during the ocean experiments conducted on August 7th and 9th, 2018. b) The SPOT Trace units attached to the top of a CODE/DAVIS surface drifter for position tracking.}
    \label{fig:SPOT_Trace_GPS}
\end{figure}

The SPOT Trace GPS units deployed during ocean experiments were used to track trajectories of surface drifters and OSCAR Manikins. This GPS system provides position measurements every 5 minutes (i.e., a sampling rate of $0.0033$~Hz) when  there is an unobstructed view of the sky. Position measurements registered by the SPOT Trace GPS are transmitted over cell phone tower networks as a text message using standardized communication protocols for short message service (SMS). Size-wise, as shown in Figure~\ref{fig:SPOT_Trace_GPS}, this GPS unit has a width of 5.13 cm, a length of 8.83 cm, and thickness of 2.14 cm. All together, these qualities have made this sensor an attractive solution to tracking position of surface drifters during ocean deployments.

\subsection{Multirotor sUAS Wind Measurements}
\label{Multiroto_UAS_wind_measurements}

\subsubsection{Quadrotor Platform}
\label{sss:quadrotor_platform}

\begin{figure}[h!]
\centering
\begin{subfigure}[b]{0.47\textwidth}
         \centering
         \includegraphics[width=\textwidth]{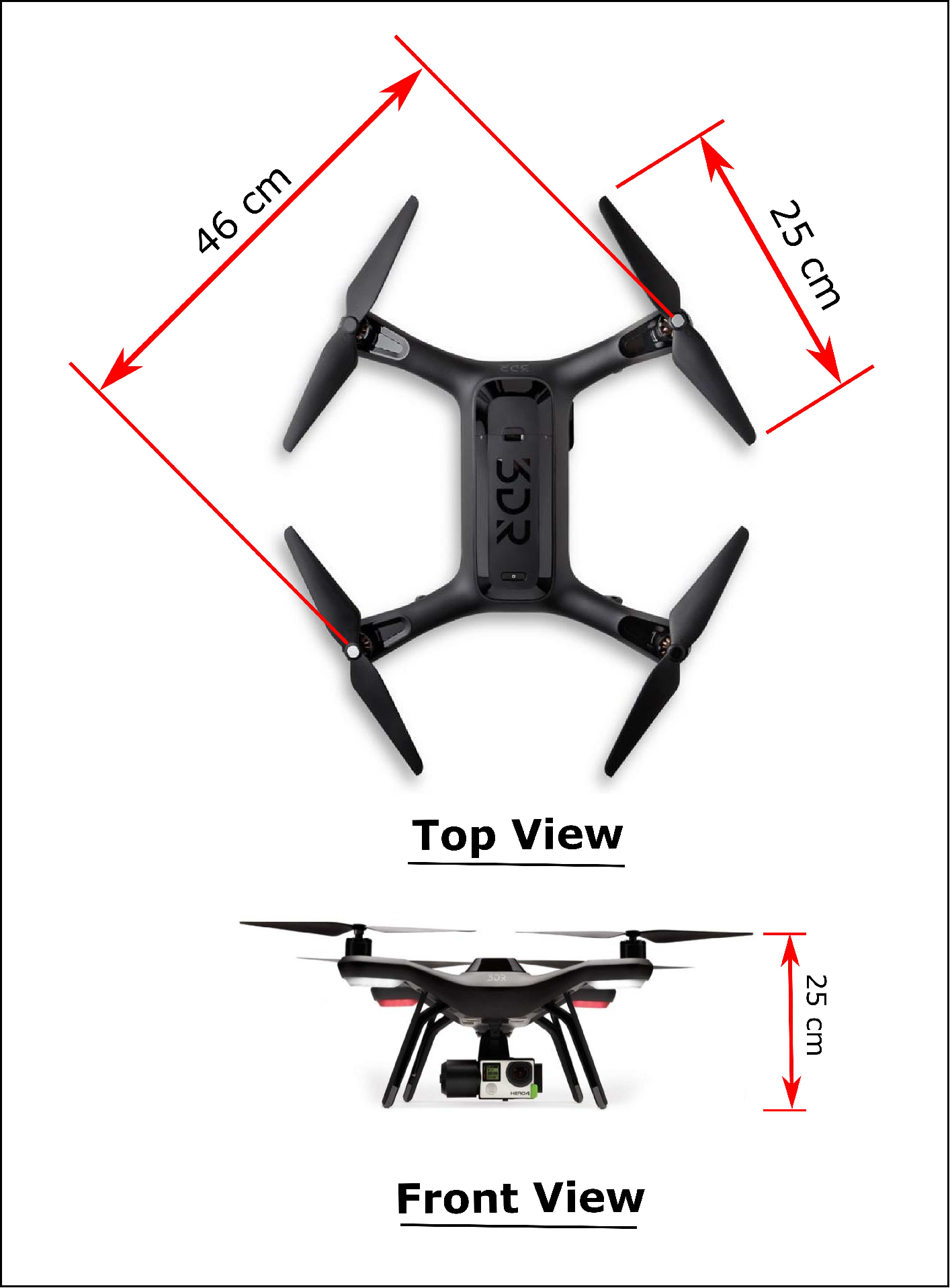}
         \caption{}
         \label{subfig:quadrotor_dimensions}
\end{subfigure}
\begin{subfigure}[b]{0.47\textwidth}
         \centering
         \includegraphics[width=\textwidth]{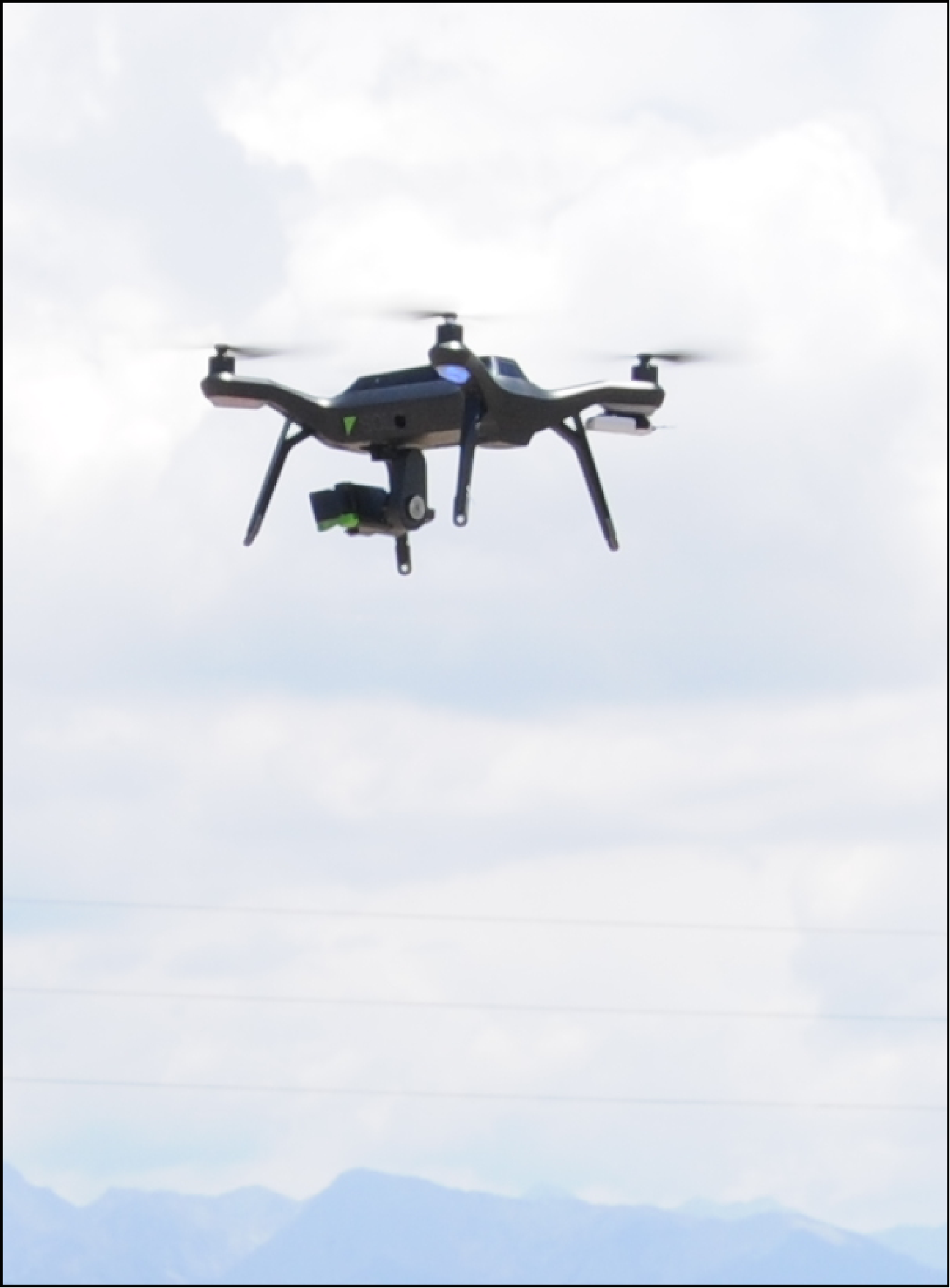}
         \caption{}
         \label{subfig:3DRSolo_Flying}
\end{subfigure}
    \caption{ a) A schematic with dimensions of the multirotor sUAS employed to measure wind velocity at 10 m ASL. }
    \label{fig:quadrotor_dimensions}
\end{figure}

The multirotor aircraft used for wind sensing is an off-the-shelf 3DR Solo quadrotor with camera and 3-axis gimbal. The 3DR Solo, as shown in Figure~\ref{fig:quadrotor_dimensions}, is 25 cm tall and has a diagonal span of 46 cm. Fully integrated with gimbal, camera, and battery the quadrotor weighs 1.5~kg and has a flight endurance of approximately 15 minutes in pristine atmospheric conditions. The quadrotor is propelled by four brushless 880 $\rm K_v$ electric motors and 25 cm x 11.4 cm self-tightening propellers. The flight controller on board with quadrotor is a Pixhawk 2.1 Green Cube, which has a boat mode feature for operating from a moving platform.

\section{Leeway Experiments Results}
\label{s:Leeway_Experiments_Results}

We conducted lake and ocean experiments employing a multirotor sUAS as well as built-in-house and off-the-shelf GPS tracking modules to characterize the leeway of a person in water. The quadrotor described in Section~\ref{sss:quadrotor_platform} was used to collect ten-minute wind velocity observations at 10~m ASL. multirotor sUAS wind observations were then used to approximate wind conditions for the duration of leeway experiments. The Pixhawk-GPS modules, built from sUAS-grade attitude and heading reference systems and GPS receivers, were used along with off-the-shelf SPOT Trace GPS units to track the position of manikins and surface drifters. Additionally, Pixhawk-GPS modules were used to measure the orientation, as well as the rotational and translational motion of every manikin. These measurements were then processed to determine the transitional and rotational displacements of each manikin relative to downwind and crosswind components. From these results we were able to estimate downwind and crosswind leeway parameters.

\subsection{Wind Measurements}
\label{ss:wind_measurements}

multirotor sUAS wind velocity observations were used to determine the prevailing wind conditions for the time lapse of lake and ocean leeway experiments. In this process, the east and north components of the wind velocity vector $\mathbf{u}_w$ (i.e., $\bm{u}$ and $\bm{v}$) were averaged over the duration of each flight. The number of averaged measurements that were collected during each experiment varied based on weather minimums for safe multirotor sUAS operations. Table~\ref{table:mutirotor_wvel} summarizes the averaged wind measurements collected during each experiment. The averaged wind velocity components $\overline{\bm{u}}$ and $\overline{\bm{v}}$ were then used to characterize a first-order fit to approximate wind conditions for the duration of each manikin release. 

\begin{figure}[h!]
\centering
\begin{subfigure}[b]{0.47\textwidth}
         \centering
         \includegraphics[width=\textwidth]{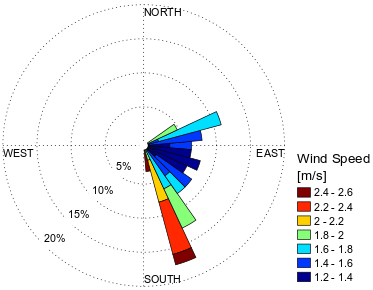}
         \caption{}
         \label{subfig:windrose25}
\end{subfigure}
\begin{subfigure}[b]{0.47\textwidth}
         \centering
         \includegraphics[width=\textwidth]{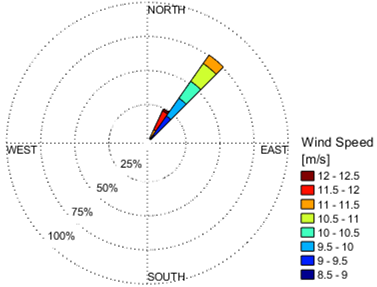}
         \caption{}
         \label{subfig:windrose7}
\end{subfigure}
\begin{subfigure}[b]{0.47\textwidth}
         \centering
         \includegraphics[width=\textwidth]{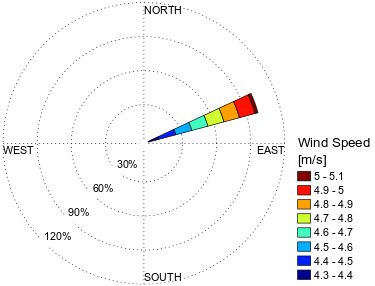}
         \caption{}
         \label{subfig:windrose9}
\end{subfigure}
  \caption{ a) Wind speed and downwind direction measured during leeway experiments in Claytor Lake, VA  on June 25th, 2018 from 9:59 to 11:03~EDT. b) Wind speed and direction measured during leeway experiments offshore from Martha's Vineyard on August 7th, 2018 from 13:58 to 15:58~EDT. c) Wind speed and direction measured during leeway experiments offshore from Martha's Vineyard on August 9th, 2018 from 10:33 to 14:12~EDT.}\label{fig:wind_rose}
\end{figure}

Analysis of wind conditions demonstrated low, moderate, and high wind speeds across all three leeway experiments. As shown in Figure~\ref{subfig:windrose25}, the downwind conditions during lake experiments conducted on June, 25th, 2018 varied between the northeast and southeast directions with maximum speed of 2.8 m/s. During the ocean experiment conducted on August 7th, 2018, the downwind conditions were along the northwest direction with a maximum speed of 12.5~m/s (see Figure~\ref{subfig:windrose7}). Finally, as shown in Figure~\ref{subfig:windrose9}, downwind wind conditions were along the east-northeast direction with maximum speeds reaching 5.1~m/s during ocean field experiments conducted on August 9th, 2018.

Additional experiments were also conducted to validate quadrotor wind estimates next to conventional in-situ and remote-sensing wind sensors at 10 m above ground level. Validation results show quadrotor wind speed estimates to have an average mean error of 0.3 m/s and 1.0 m/s relative to wind speed observations collected with in-situ and remote-sensing instruments, respectively. The average mean error of quadrotor wind direction estimates was measured as 9.9$^\circ$ compared to observations from in-situ and remote sensors. Additional details and results of validation experiments are presented in Appendix~\ref{as:quadrotor_wind_estimation_framework}.

\begin{table}[htbp]
\centering
  \caption{Averaged multirotor sUAS wind velocity measurements collected during lake and ocean leeway experiments in units of m/s. }

\resizebox{\textwidth}{!}{%
  \begin{tabular}{ccccccccccccccc}
    \toprule
   
      \multicolumn{4}{c}{ Lake Experiments 06/25/2018 }&& \multicolumn{4}{c}{ Ocean Experiments 08/07/2018 } &&\multicolumn{4}{c}{ Ocean Experiments 08/09/2018 }\\
      \cline{1-4} \cline{6-9} \cline{11-14} \\[-1em]
        \rule{0pt}{10pt}Time UTC & $\overline{\bm{u}}$& $\overline{\bm{v}}$ & $\overline{U}$& & Time UTC & $\overline{\bm{u}}$& $\overline{\bm{v}}$&$\overline{U}$& & Time UTC & $\overline{\bm{u}}$& $\overline{\bm{v}}$& $\overline{U}$ \\
      \toprule \midrule 
  14:26 & 1.0 & -0.9 & 1.3 && 18:43&6.1 & 8.4 & 10.4 &&15:53&5.6&0.7&5.6\\ 
  14:41 & 1.4 & -0.5 & 1.5 && 19:14& 7.1 &7.0& 10.0&&16:08&4.3&1.9&4.7\\  
  14:54 & 1.5 & 0.7  & 1.6 && 19:29& 6.0 & 10.5&12.1 &&17:13&3.9&1.8&4.3\\
  --    & --  & --   & --&& -- & -- & -- & -- & & 17:58 & 4.4& 2.5&5.0\\
  --    & --  & --   & --&& -- & -- & -- & -- & & 18:08 & 4.6& 0.6&4.6\\

  \midrule
  \bottomrule
  \end{tabular}}
  \label{table:mutirotor_wvel}
  \end{table}

\subsection{Position Tracking}
\label{ss:position_tracking}

The global position of manikin-and-drifter pairs was tracked during leeway experiments using Pixhawk-GPS modules and SPOT Trace GPS units in varied configurations. During the Claytor Lake leeway experiment, Pixhawk-GPS modules alone were used to track a single manikin-and-drifter pair. Alternatively, Pixhawk-GPS modules were used during ocean experiments to track manikins only. Surface drifters released into the ocean along with manikins were tracked using SPOT Trace GPS units instead. Resolution discrepancies across the two GPS systems were reconciled by smoothing Pixhawk GPS measurements to match the temporal resolution of SPOT Trace GPS units using a 5-minute moving average. The accuracy of both GPS systems was also assessed using the validation analysis described in~\ref{s:GPS_performance_comparison}. Results from this assessment demonstrate that in addition to a higher resolution, Pixhawk GPS units also have a greater accuracy.

\begin{figure}[h!]
\centering
\begin{subfigure}[b]{0.47\textwidth}
         \centering
         \includegraphics[width=\textwidth]{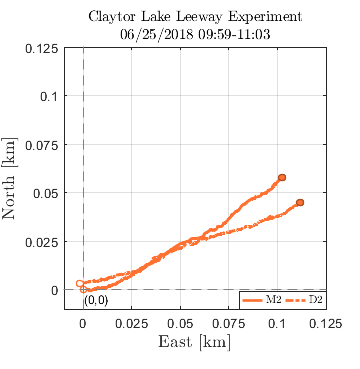}
         \caption{}
         \label{subfig:drifter_manikin_position_625A}
\end{subfigure}
\begin{subfigure}[b]{0.47\textwidth}
         \centering
         \includegraphics[width=\textwidth]{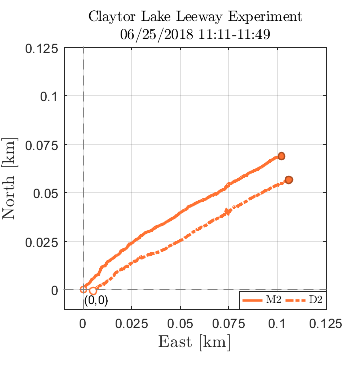}
         \caption{}
         \label{subfig:drifter_manikin_position_625B}
\end{subfigure}
 \caption{The global position tracking of a manikin-and-drifter pair released in Claytor Lake, VA on June 25th, 2018 from a) 9:59 to 11:03 and b) 11:11 to 11:49 EDT}\label{fig:DM_lake_position}
\end{figure}

A single manikin-and-drifter pair was released during the leeway experiments conducted in Claytor Lake, VA on June 25th, 2018. Figures~\ref{subfig:drifter_manikin_position_625A} and \ref{subfig:drifter_manikin_position_625B} show the drifter-and-manikin release broken into two segments (i.e., 1A and 1B) extending from 9:59 to 11:03 EDT and 11:11 to 11:49~EDT. The total displacement of the manikin during the first and second segments reached 0.24~km and 0.18~km, respectively (see Table~\ref{table:tracking7_9}). Across both segments, the maximum separation between drifter and manikin was measured to be 0.02~km and 0.01~km for segments 1 and 2 respectively. 

\begin{figure}[h!]
\centering
\begin{subfigure}[b]{0.47\textwidth}
         \centering
         \includegraphics[width=\textwidth]{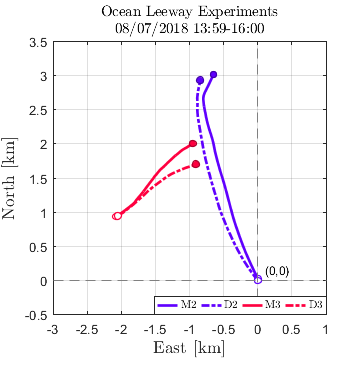}
         \caption{}
         \label{subfig:drifter_manikin_position_87}
\end{subfigure}
\begin{subfigure}[b]{0.47\textwidth}
         \centering
         \includegraphics[width=\textwidth]{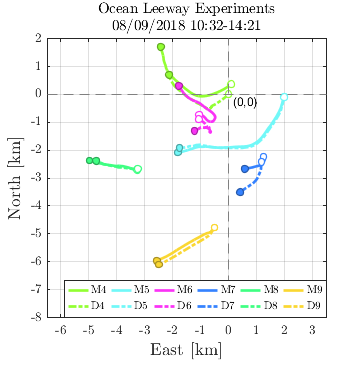}
         \caption{}
         \label{subfig:drifter_manikin_position_89}
\end{subfigure}
 \caption{a) Global position tracking of two manikin-and-drifter pairs released into the ocean southwest of Martha's Vineyard, MA on August 7th, 2018. b) Global position tracking of six manikin-and-drifter pairs released into the ocean southeast of Martha's Vineyard, MA on August 9th, 2018.}\label{fig:DM_position}
\end{figure}

In total, eight manikin-and-drifter releases were tracked during the ocean leeway experiments. As shown in Figure~\ref{subfig:drifter_manikin_position_87}, two manikin-and-drifter releases were conducted on August 7th, 2018 from 17:58 to 19:58 UTC. Figure~\ref{subfig:drifter_manikin_position_89} shows the six releases that were performed on August 9th, 2018 from 14:33 to 18:12 UTC. Varied drift characteristics were observed across both sets of ocean releases. On August 7th, 2018, Manikin~2 (i.e., M2) traveled 3.07~km away from its release point while Manikin~3 only traveled 1.55~km over a period of an hour and fifty-nine minutes (see Table~\ref{table:tracking7_9}). During the  August 9th, 2018 experiments, the total displacement of manikins ranged from 0.8 km to 4.21~km over a time period of five hours and forty-nine minutes. Additionally, as shown in Table~\ref{table:tracking7_9}, manikin-and-drifter separation distance  varied across all nine releases from 0.16~km to 1.70~km. This information combined with the accuracy GPS measurements can help identify the sources of error for leeway parameter estimates.

\begin{table}[tbh!]\small
\centering
  \caption{Summary of the total displacement of manikins and the maximum separation between manikins and surface drifters during Claytor Lake and ocean leeway experiments.}
  \resizebox{\textwidth}{!}{%
  \begin{tabular}{ccccccccccccccc}
    \toprule
      Manikin & Date&  Duration [hrs:min]& Total Distance [km] & Maximum Separation [km]\\
      \toprule \midrule
  1A &Jun. 25th, 2018 & 1:04 & 0.24 & 0.02 \\ 
  1B &Jun. 25th, 2018 & 0:38 & 0.18 & 0.01 \\ 
  2 &Aug. 7th, 2018 & 1:59 & 1.55 & 0.31 \\ 
  3 &Aug. 7th, 2018 & 1:59 & 3.07& 0.21  \\  
  4 &Aug. 9th, 2018 & 3:49 & 2.87 & 1.05 \\
  5 &Aug. 9th, 2018 & 3:49 & 4.31 & 0.19\\
  6 &Aug. 9th, 2018 & 3:49 & 1.26 & 1.70   \\
  7 &Aug. 9th, 2018 & 3:49 & 0.80 & 0.85\\
  8 &Aug. 9th, 2018 & 3:49 & 1.51 & 0.24 \\
  9 &Aug. 9th, 2018 & 3:49 & 2.39 & 0.16\\
 
  \midrule
  \bottomrule
  \end{tabular}}
  \label{table:tracking7_9}
  \end{table}

\subsection{Leeway Estimates}
\label{ss:leeway_estimates}

\begin{figure}[h!]
\centering
\begin{subfigure}[b]{0.47\textwidth}
         \centering
         \includegraphics[width=\textwidth]{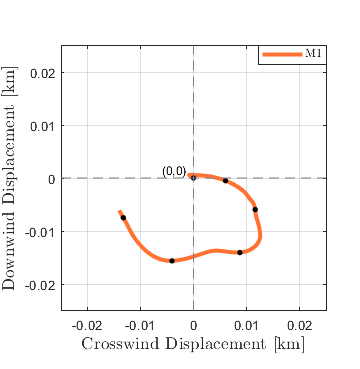}
         \caption{}
         \label{subfig:PVD_625A}
\end{subfigure}
\begin{subfigure}[b]{0.47\textwidth}
         \centering
         \includegraphics[width=\textwidth]{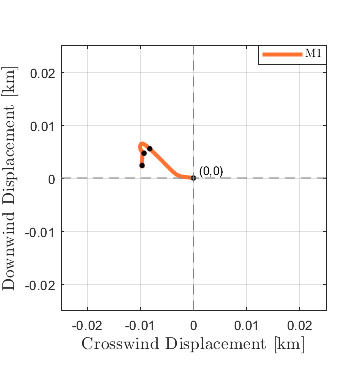}
         \caption{}
         \label{subfig:PVD_625B}
\end{subfigure}
\caption{The progressive vector diagrams for releases conducted in Claytor Lake, VA on June 25th, 2018 from a) 9:59 - 11:03~EDT and b) 11:11-11:49~EDT.  }\label{fig:Lake_PVD}
\end{figure}

\begin{figure}[h!]
\centering
\begin{subfigure}[b]{0.47\textwidth}
         \centering
         \includegraphics[width=\textwidth]{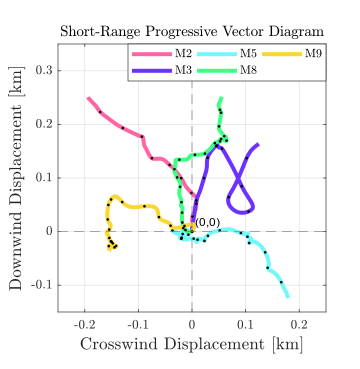}
         \caption{}
         \label{subfig:PVD-SR}
\end{subfigure}
\begin{subfigure}[b]{0.47\textwidth}
         \centering
         \includegraphics[width=\textwidth]{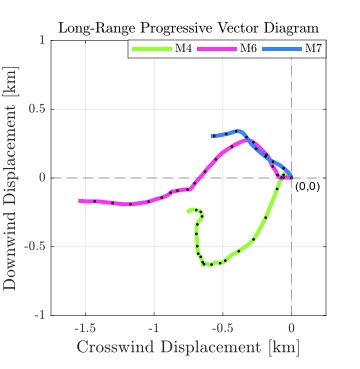}
         \caption{}
         \label{subfig:PVD-LR}
\end{subfigure}
    \caption{ a) The progressive vector diagram showing short-rage displacements that do not exceed 0.5~km.  b) The progressive vector diagram showing long-rage displacements that exceed 0.5~km.}\label{fig:PVD}
\end{figure}

Multirotor wind velocity observations and GPS position information were used to determine the downwind and crosswind flow-relative displacement of each manikin. As in previous studies (i.e., \cite{breivik2011wind,allen2005leeway}), this information was utilized to generate progressive vector diagrams (PVDs) to assess the downwind and upwind displacement divergence and jibing behavior of all nine manikins. Results from this analysis are presented in Figures~\ref{subfig:PVD_625A} and \ref{subfig:PVD_625B} for the manikin-and-drifter release conducted in Claytor Lake.  PVDs for ocean manikin releases are shown in Figure~\ref{subfig:PVD-SR} for manikins whose total displacement did not exceed 0.5~km and in Figure~\ref{subfig:PVD-LR} for manikins whose total displacement was greater than 0.5~km. Significant variability was observed in the divergence of upwind and downwind displacement of all nine manikins. Almost all of the manikins exhibited left and right divergence for downwind and upwind displacements at separate times. Additionally, based on the spatial separation between five-minute position averages, different periods of low and high displacement rates were observed for each manikin. The frequency and scale of jibes observed also varied throughout the drifting period of each manikin.

\begin{figure}[h!]
\centering
\begin{subfigure}[b]{0.47\textwidth}
         \centering
         \includegraphics[width=\textwidth]{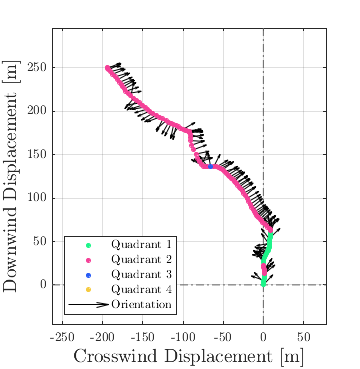}
         \caption{}
         \label{subfig:PVD2}
\end{subfigure}
\begin{subfigure}[b]{0.47\textwidth}
         \centering
         \includegraphics[width=\textwidth]{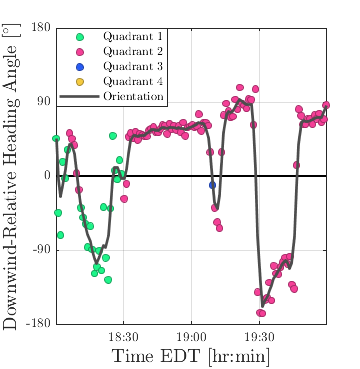}
         \caption{}
         \label{subfig:drifter_manikin_downwind-relative_orientation87}
\end{subfigure}
    \caption{ a) The progressive vector diagram and downwind-relative orientation of Manikin 2.  b) A time series of the downwind-relative orientation of Manikin 2.}\label{fig:DM2}
\end{figure}

\begin{figure}[h!]
\centering
\begin{subfigure}[b]{0.47\textwidth}
         \centering
         \includegraphics[width=\textwidth]{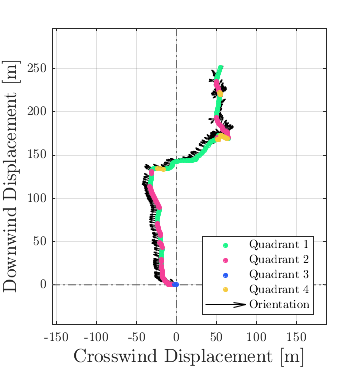}
         \caption{}
         \label{subfig:PVD8}
\end{subfigure}
\begin{subfigure}[b]{0.47\textwidth}
         \centering
         \includegraphics[width=\textwidth]{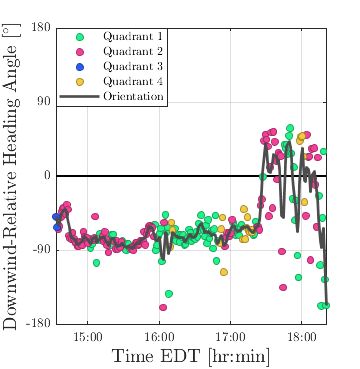}
         \caption{}
         \label{subfig:drifter_manikin_downwind-relative_orientation89}
\end{subfigure}

   \caption{ a) The progressive vector diagram and downwind-relative orientation of Manikin 8.  b) A time series of the downwind-relative orientation of Manikin 8.}\label{fig:DM8}
\end{figure}

 Information from the attitude and heading reference system inside of each Pixhawk-GPS module were also used to determine the downwind-relative orientation of manikins. These measurements, which are not available with conventional GPS systems, can help understand how the drift characteristics of each manikin are influenced by downwind-relative orientation. For example, the PVDs shown in Figures~\ref{subfig:PVD2} and~\ref{subfig:PVD8} combine the positive and negative divergence patterns of downwind and upwind displacements with measurements of downwind-relative orientation for Manikins~2 and 8, respectively. From this analysis we observe that there were periods of drifting where changes in downwind-relative orientation and jibing behavior were significantly correlated. Moreover, from each time series of downwind-relative orientation we see that manikins have a tendency to converge to some orientations more than others. 

\subsection{Leeway Parameter Estimates}
\label{ss:leeway_parameter_estimates}

\begin{figure}[h!]
\centering
\begin{subfigure}[b]{0.47\textwidth}
         \centering
         \includegraphics[width=\textwidth]{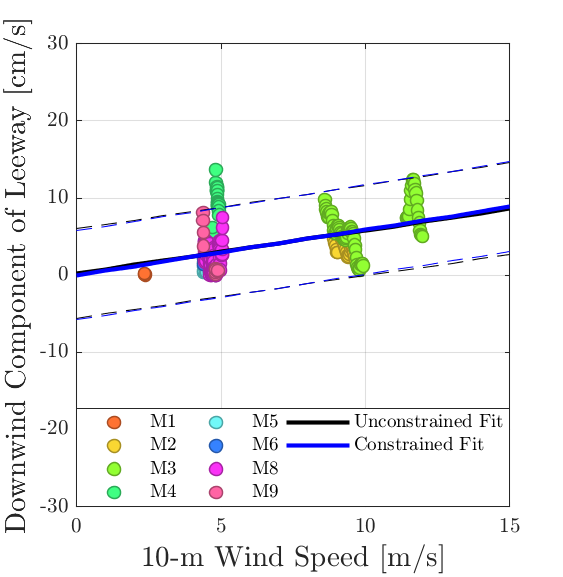}
         \caption{}
         \label{subfig:dcl_pp}
\end{subfigure}
\begin{subfigure}[b]{0.47\textwidth}
         \centering
         \includegraphics[width=\textwidth]{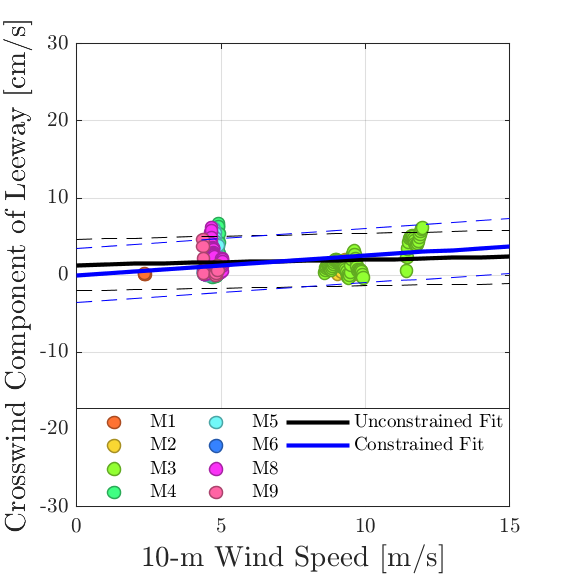}
         \caption{}
         \label{subfig:ccl_pp}
\end{subfigure}
    \caption{a) First quadrant downwind component of leeway. b) First quadrant crosswind components of leeway.}\label{fig:CL_pp}
\end{figure}

\begin{figure}[h!]
\centering
\begin{subfigure}[b]{0.47\textwidth}
         \centering
         \includegraphics[width=\textwidth]{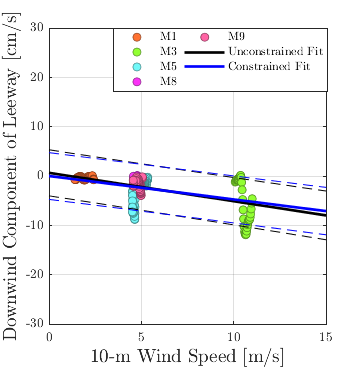}
         \caption{}
         \label{subfig:dcl_np}
\end{subfigure}
\begin{subfigure}[b]{0.47\textwidth}
         \centering
         \includegraphics[width=\textwidth]{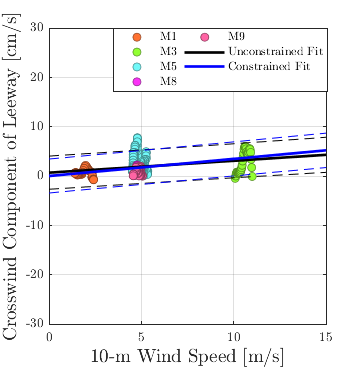}
         \caption{}
         \label{subfig:ccl_np}
\end{subfigure}
   \caption{a) Second quadrant downwind component of leeway. b) Second quadrant crosswind components of leeway.}\label{fig:CL_np}
\end{figure}

\begin{figure}[h!]
\centering
\begin{subfigure}[b]{0.47\textwidth}
         \centering
         \includegraphics[width=\textwidth]{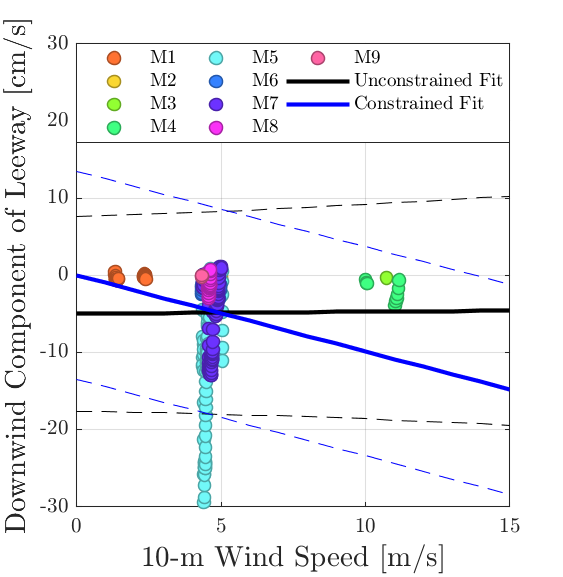}
         \caption{}
         \label{subfig:dcl_nn}
\end{subfigure}
\begin{subfigure}[b]{0.47\textwidth}
         \centering
         \includegraphics[width=\textwidth]{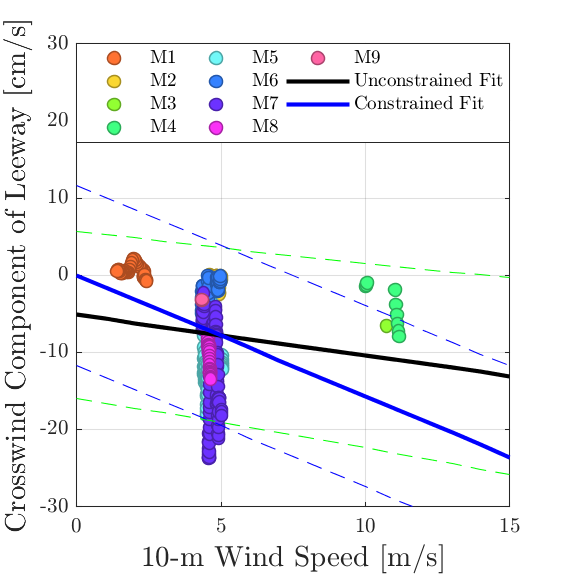}
         \caption{}
         \label{subfig:ccl_nn}
\end{subfigure}
  \caption{a) Third quadrant downwind component of leeway. b) Third quadrant crosswind components of leeway.  }\label{fig:CL_nn}
\end{figure}

\begin{figure}[h!]
\centering
\begin{subfigure}[b]{0.47\textwidth}
         \centering
         \includegraphics[width=\textwidth]{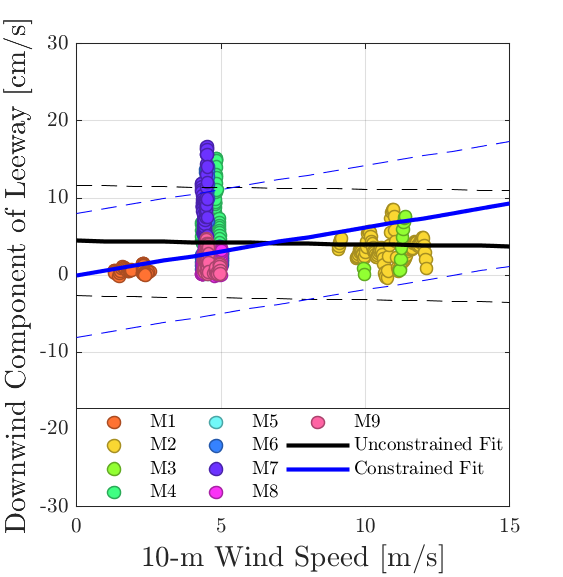}
         \caption{}
         \label{subfig:dcl_pn}
\end{subfigure}
\begin{subfigure}[b]{0.47\textwidth}
         \centering
         \includegraphics[width=\textwidth]{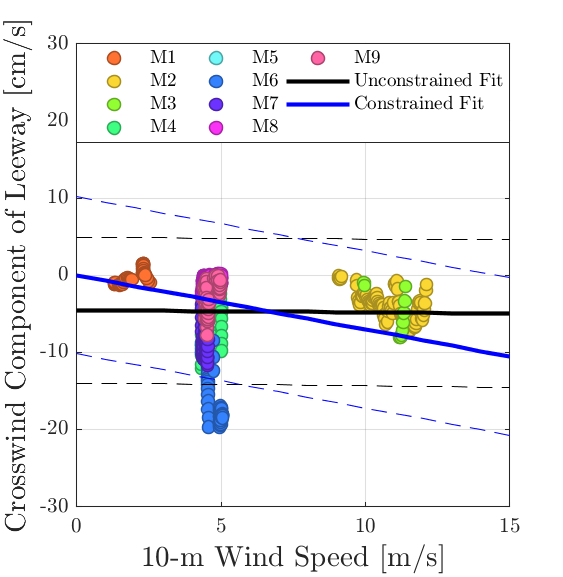}
         \caption{}
         \label{subfig:ccl_pn}
\end{subfigure}
  \caption{a) Fourth quadrant downwind component of leeway. b) Fourth quadrant crosswind components of leeway.  }\label{fig:CL_pn}
\end{figure}

\begin{figure}[h!]
\centering
\begin{subfigure}[b]{0.47\textwidth}
         \centering
         \includegraphics[width=\textwidth]{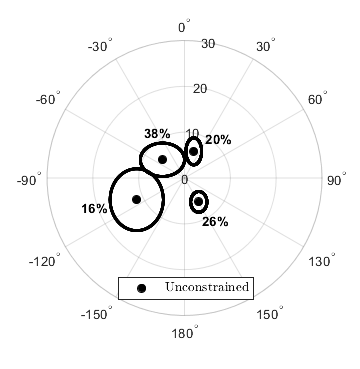}
         \caption{}
         \label{subfig:leeway_values_unconstrained}
\end{subfigure}
\begin{subfigure}[b]{0.47\textwidth}
         \centering
         \includegraphics[width=\textwidth]{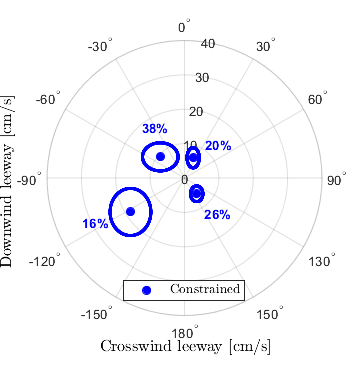}
         \caption{}
         \label{subfig:leeway_values_constrained}
\end{subfigure}
   \caption{ Distribution of measured manikin leeway values, in each of the four quadrants in the two-dimensional leeway-space. 
     The downwind (vertical axis) and crosswind (horizontal axis) leeway (measured in cm/s) for a hypothetical 10-m height wind speed of 10 m/s. The distance from the origin indicates leeway speed (cm/s) while the angle relative to the vertical indicates the object’s divergence from the wind direction (wind blowing upwards along the vertical axis).
     The ellipses show crosswind and downwind error.
     The values for manikins are shown and were studied using the indirect method.
    Overall, there is slight bias toward positive downwind and negative crosswind leeway, as also seen for persons-in-water in \cite{breivik2011wind}. 
   Subfigures a) and b) show leeway values  estimated using the unconstrained  and constrained regression moldes, respectively.
    }\label{fig:leeway_values}
\end{figure}

Wind velocity and global positioning information of manikins and accompanying drifters were also used to characterize leeway parameters following the method described in Section~\ref{ss:windage_analysis}. In this process, downwind and crosswind leeway models were characterized as a function of the 10-m wind speed using constrained and unconstrained linear regression (see Figures~\ref{fig:CL_pp}-\ref{fig:CL_pn}). Results from constrained and unconstrained linear regression, which include slope, offset values and error bounds, are shown in Tables~\ref{table:leeway_components_constrained} and \ref{table:leeway_components_unconstrained}, respectively. The downwind and crosswind models were used to assess the leeway at 10~m/s as shown in Figures~\ref{subfig:leeway_values_unconstrained} and \ref{subfig:leeway_values_constrained}. Additionally, leeway speed and divergence at wind speeds of 10~m/s were plotted with error bounds included for the constrained and unconstrained cases as shown in Figures~\ref{subfig:leeway_values_constrained} and \ref{subfig:leeway_values_unconstrained}. From this analysis, we see the trends in divergence for observations corresponding to one of four quadrants in the two-dimensional leeway space. Percentage-wise, the leeway for a person in water fell in quadrant 1 twenty percent of the time, in quadrant 2 thirty eight percent of the time, in quadrant 3  sixteen percent of the time and in quadrant 4 twenty six percent of the time.

\begin{table}[tbh!]\small
\centering
  \caption{Summary of downwind and crosswind leeway components estimated from unconstrained linear regression.}

  \begin{tabular}{ccccccccccccccc}
    \toprule
    \multirow{2}{*}{Quadrant}&
      \multicolumn{3}{c}{ Downwind Leeway }&& \multicolumn{3}{c}{ Crosswind Leeway }\\ 
      \cline{2-4} \cline{6-8}
        \rule{0pt}{10pt} & $C_w$ [$\%$] & $b_w$ [cm/s]& $\varepsilon_w$ [cm/s]& &$C_w^\perp$ [$\%$] & $b_w^\perp$ [cm/s]& $\varepsilon_w^{\perp}$ [cm/s]\\
      \toprule \midrule
  1 & 0.56 & 0.23 &2.94 && 0.07 & 1.32 & 1.72\\ 
  2 &-0.05 & 4.47 &3.64 &&-0.03 & 4.47 & 4.82 \\ 
  3 &0.03 & -5.0  &6.75 &&-0.53 & -5.14 & 5.80\\
  4 &-0.62 & 1.02 &2.26 && 0.24 & 1.02 &1.77 \\
  \midrule
  \bottomrule
  \end{tabular}
  \label{table:leeway_components_unconstrained}
  \end{table}

\begin{table}[tbh!]\small
\centering
  \caption{Summary of downwind and crosswind leeway components estimated from constrained linear regression.}

  \begin{tabular}{ccccccccccccccc}
    \toprule
    \multirow{2}{*}{Quadrant}&
      \multicolumn{3}{c}{ Downwind Leeway }&& \multicolumn{3}{c}{ Crosswind Leeway }\\ 
      \cline{2-4} \cline{6-8}
        \rule{0pt}{10pt} & $C_w$ [$\%$] & $b_w$ [cm/s]& $\varepsilon_w$ [cm/s]& & $C_w^\perp$ [$\%$] & $b_w^\perp$ [cm/s]& $\varepsilon_w^\perp$ [cm/s]\\
      \toprule \midrule
  1 & 0.59 & -- & 2.94 && 0.25&--& 1.78\\ 
  2 & 0.62 & -- & 4.09 &&-0.70&--& 5.17 \\  
  3 &-0.99 & -- & 6.85 &&-1.57&--& 5.94\\
  4 &-0.46 & -- & 2.29 && 0.35&--& 1.79 \\
  \midrule
  \bottomrule
  \end{tabular}
  \label{table:leeway_components_constrained}
  \end{table}

\section{Discussion}
\label{s:Discussion}

Nine manikin-and-drifter pairs were released during ocean and lake experiments to determine the leeway of a person in water. A single release was conducted in Claytor Lake, Virginia and eight others in the Atlantic Ocean south of Martha's Vineyard, Massachusetts. In all three leeway experiments, a quadrotor was  deployed to measure the wind in ten-minute intervals at 10 m ASL. Interpolation of ten-minute averages of multirotor wind estimates showed the wind conditions to range between 1.2~m/s and 2.6~m/s during lake experiments and between 4.3~m/s and 12.5~m/s during ocean experiments. Across all nine manikin-and-drifter releases, seven were conducted in wind conditions for which the performance of quadrotor wind estimation is well validated, i.e., $\mathbf{u}_w<$6~m/s (see Appendix~\ref{ass:wind_estimation_experimental_validation}). Moreover, the higher wind conditions observed during ocean experiments were well above the typical 10-m wind speed average of 5~m/s recorded at the Martha's Vineyard Airport. Therefore, although there is a need to test the reliability of multirotor wind measurements at higher wind speeds, the validated performance is well within typical wind conditions expected for Martha's Vineyard.

Attitude and heading reference system modules (Pixhawk-GPS) and SPOT Trace GPS antennae units were used during leeway experiments to track the drifting trajectories of manikins and drifters. Significant variability was observed across the displacement of  manikin-and-drifter pairs. During the lake experiments, the maximum distanced traveled by manikin experiments ranged between 0.18~km and 0.24~km. In the ocean, manikin displacements ranged between 1.55~km and 3.07~km on the first day and between 1.26~km and 4.31~km on the second day. Furthermore, the separation between manikins and drifters also varied considerably during ocean experiments compared to lake experiments. The maximum separation between a drifter and its paired manikin  was measured as 0.02~km in Claytor Lake and 1.70~km in the ocean. Moreover, significant divergence between manikin and surface drifter trajectories were observed in the ocean within a distance of 0.25~km (i.e. Manikin 5 In Figure~\ref{subfig:drifter_manikin_position_89}). The difference in manikin and drifter displacements across all nine releases is attributed to the difference in surface flow uniformity across lake and ocean environments. 

Wind velocity and global position information was used to resolve the flow-relative displacement of each manikin with respect to downwind and crosswind components. The flow-relative displacement of manikins fell into all four quadrants of the two-dimensional wind-relative reference frame (i.e., the leeway space, Figure \ref{fig:leeway_polar_diagram}). 
This result suggests that the manikins experienced both downwind and upwind displacements, which could be genuine, suggesting further investigation, possibly  the result of wind-relative orientation effects. Another possible reason for the negative downwind leeway values measures is vertical shear (in the lake) or too large of a separation of the manikin from the drifter (in the ocean, that is, the manikin experienced a different local current than the drifter), or the effect of waves. 

Significant spread in downwind and crosswind leeway measurements was also observed for wind speeds of 5~m/s (see Figures~\ref{subfig:dcl_pp}-\ref{subfig:ccl_nn}). Considering that the validation experiments described in Appendix~\ref{ass:wind_estimation_experimental_validation} demonstrate multirotor wind estimates to be reliable below 6~m/s, the spread in downwind and crosswind leeway components is attributed to limitations of the indirect method measuring flow-relative velocity. A review presented in~\cite{breivik2011wind} of direct and indirect methods for measuring flow-relative velocity notes that the indirect method is prone to error as drifter and manikin pairs drift apart. Additionally, analysis described in~\ref{as:gps_perfromance_assesment} comparing GPS position observations show the SPOT Trace GPS observations to have on average a RMSE of 21~m relative to Pixhawk-GPS measurements. Therefore, error in SPOT Trace GPS position measurements, which are differentiated over over five-minute intervals to obtain flow-relative velocity measurements, can also contribute to the spread of downwind and crosswind leeway observations.

Flow-relative displacements falling into one of four quadrants of the wind reference frame and wind speed measurements were used to characterize four sets of downwind and crosswind leeway models. In this process, the leeway parameters, and the associated offset and error values, were estimated using constrained and unconstrained linear regression (see Figure \ref{fig:leeway_values} and Tables \ref{table:leeway_components_unconstrained} and \ref{table:leeway_components_constrained}). A significant subset of leeway parameters were found to be close proximity to person-in-water leeway values reported in Figure 1 of~\cite{breivik2011wind} based on experiments described in \cite{allen2005leeway}.

The application of multirotor sUAS during leeway experiments provided on-demand 10-m height wind observations that otherwise would be challenging to attain using conventional wind sensors. The small footprint required to launch, operate and recover the multirotors  made deployment from a small vessel possible. Additionally, the capability to collect wind observations hovering at 10~m ASL circumvented the need to extrapolate observations as required in situations where wind sensors are located just centimeters above the surface. We do note that the short endurance of the multirotor sUAS did present some limitations for safe operations.
 
The attitude and heading reference system modules enabled new capabilities for monitoring drifting manikins. In addition to high-resolution and accurate position tracking, these modules provide attitude and heading data, which allow for analysis that is not possible with conventional methods that provide only coarsely sampled position. High resolution observation of position and orientation allow for small-scale measurements that cannot be resolved with coarse measurements from conventional GPS sensors. This new capability can aid in potentially new forecasting models which account for rotational dynamics as well.

We found a number of limitations that need to be addressed in future studies to increase the confidence in leeway parameters obtained via the method employed in this study. Multirotor wind observations were limited to 10 minutes, due to battery life. Additionally, wind estimates need to be validated for the entire range of wind conditions higher than 4~m/s. Additional experiments are required next to an independent sensor to characterize error in wind estimates over a wide range of velocities. Future experiments also need to employ flowmeters for surface current observations as it is possible that significant leeway error may be the result of non-uniform flow, that is, the manikin experiences a slightly different flow field than its accompanying surface drifter, especially as they drift farther form one another. Moreover, the last day may be over-sampled in comparison with the entire ensemble of data and may not be representative of day-to-day conditions. 

Overall, results from this study demonstrate the potential that multirotor sUAS and attitude and heading reference system modules have to  improve measurements with applications to forecasting for search-and-rescue. On-demand multirotor sUAS wind measurements can circumvent the need to integrate a wind sensor onto a drifting object which may affect drifting characteristics and require extrapolation to the 10-m height. Additionally, high-resolution observations of translational and rotational motions attained from the attitude and heading reference system, GPS antenna and state estimator can help develop higher-fidelity leeway models that account for changes in downwind-relative orientation. These models can then be used to study how likely are objects to converge to some orientations versus others based on geometry. 

Finally, understanding how drifting characteristics are altered by changes in downwind-relative orientation, in addition to improving trajectory forecast models, can help develop additional technologies for search-and-rescue scenarios. For example, personal flotation devices can be developed whose shape can aid a person in distress drift toward a region where they would be more likely to be found (see, e.g., \cite{serra2020search}). Ultimately, the combined impacts enabled by multirotor sUAS and attitude and heading reference system modules can improve emergency response during search-and-rescue events.

\section{Conclusion}
\label{s:conclusion}
 Multirotor sUAS and sUAS-grade navigation technology can provide high-resolution ambient measurements to characterize the leeway of small objects. In this paper we present the application of quadrotor and tracking devices built from sUAS-grade attitude and heading reference systems and GPS antennas to obtain wind velocity and surface current measurements to characterize leeway properties of small and irregularly-shaped objects. The reliability of both instruments was tested during leeway experiments performed in lake and ocean aquatic environments as well as additional experiments. Results demonstrated that the multirotor sUAS technology can be effectively leveraged to gather wind and surface current observations needed to estimate leeway parameters. Moreover, high resolution measurements of orientation were found to provide new observations to understand how the downwind-relative orientation of manikins affects drifting characteristics (i.e., downwind displacement and jibing).
 
 Future work leveraging multirotor sUAS and derivative technology to characterize the leeway of small objects needs to address various limitations. First, to improve the duration of wind observations without reducing margins of safe operations, the multirotor sUAS employed in leeway experiments needs to have an endurance that exceeds at least 15 minutes. Second, multirotor wind estimates need to be validated next to conventional wind sensors over a higher range of wind conditions. Third, leeway experiments need to allow for shorter drifting periods to reduce drifter and manikin separation or should employ a flow meter attached to the drifting object in lieu of using a surface drifter. Together, these modifications can increase the reliability of wind velocity and surface current observations attained for leeway characterization of small objects.

 \section*{Author Contributions}
\textbf{Javier Gonz\`alez-Rocha:} Conceptualization, Methodology, Validation, Formal Analysis, Investigation, Visualization, Data Curation and and Writing- Original draft preparation. \textbf{Alejandro Sosa:} Investigation, Methodology and Validation. \textbf{Regina Hanlon:} Investigation, Validation and Writing - Review $\&$ Editing. \textbf{Arthur A. Allen:} Formal Analysis, Software and Writing - Review $\&$ Editing. \textbf{Irina I. Rypina:} Investigation, Resources and Writing - Review $\&$ Editing. \textbf{David G. Schmale III:} Investigation, Resources,Funding acquisition, Supervision and Writing - Review $\&$ Editing. \textbf{Shane D. Ross:} Conceptualization, Methodology, Validation, Formal Analysis, Investigation, Visualization, Data Curation, Funding acquisition, Supervision and Writing - Review $\&$ Editing.

 \section*{Acknowledgements} Formal Analysis
This research was supported in part by grants from the National Science Foundation (NSF) under grant number AGS 1520825 (Hazards SEES: Advanced Lagrangian Methods for Prediction, Mitigation and Response to Environmental Flow Hazards) and DMS 1821145 (Data-Driven Computation of Lagrangian Transport Structure in Realistic Flows) as well as the NASA Earth and Space Science Fellowship under grant number 80NSSC17K0375. The Authors would also like to acknowledge the Virginia Tech undergraduate students Simran Singh and Christopher Rodulfo for their help testing hardware prior to conducting field experiments. We would like to thank Michael Allshouse and Thomas Peacock for help with the experiments and stimulating discussions. Additionally, we are thankful to Nicholas Fillo, who serves as the Observing Program Leader for the National Weather Service Office in Blacksburg, VA, for coordinating access to the Virginia Tech/Montgomery Executive Airport where GPS validation experiments were conducted.

\appendix

\section{GPS Performance Assessment}
\label{as:gps_perfromance_assesment}

\subsection{Validation of Pixhawk-GPS Measurements}
 \label{validation_of_pixhawk-GPS_measurements}

\begin{figure}[h!]
\centering
\begin{subfigure}[b]{0.47\textwidth}
         \centering
         \includegraphics[width=\textwidth]{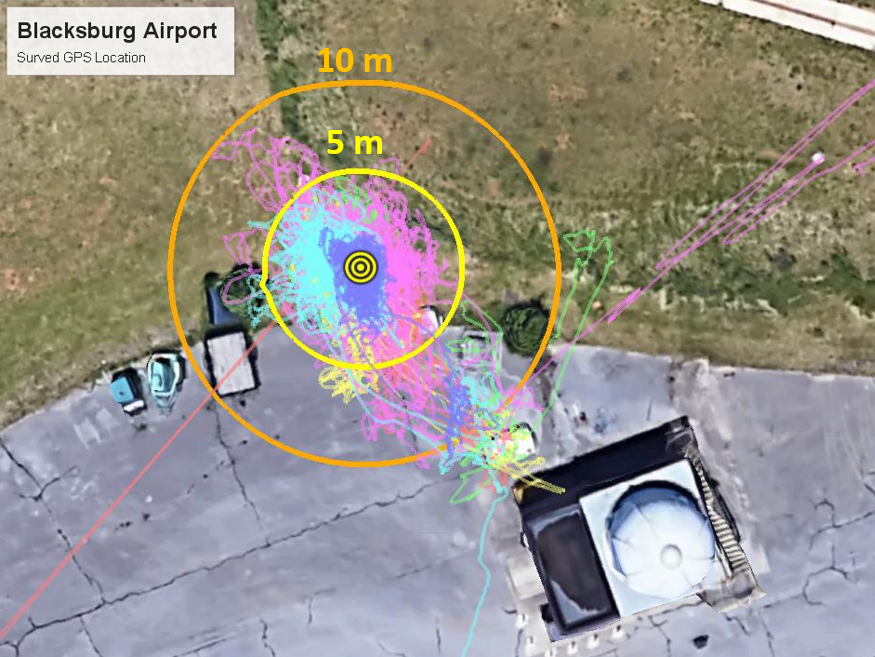}
         \caption{}
         \label{subfig:Blacksburg_surveyed_coordinates}
\end{subfigure}
\begin{subfigure}[b]{0.47\textwidth}
         \centering
         \includegraphics[width=\textwidth]{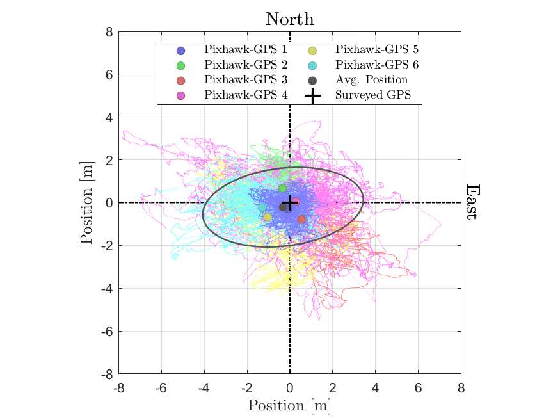}
         \caption{}
         \label{subfig:Pixhawk-GPS_error}
\end{subfigure}
   \caption{Results from field experiments to characterize the position-tracking error of Pixhawk-GPS modules. a) Top view of the surveyed marker and Pixhawk-GPS measures  with 5-m and 10-m distance radii overlaid. b) Position error measurements determined from Pixhawk-GPS position measurements recorded over the course of approximately 3 hours.}
    \label{fig:Pixhawk-GPS_validation}
\end{figure}

 To characterize the accuracy of the Pixhawk-GPS modules, field experiments were conducted next to a surveyed marker from the National Geodetic Survey.  The surveyed marker (lat 37.20601, lon -80.41452) is located within grounds of the Blacksburg Airport and was accessed with permission from the National Weather Service NWS Forecast Office located in Blacksburg, VA. During the experiment, the Pixhawk-GPS modules were turned on and then left to record position measurements laying over the surveyed marker from 20:14 to 22:56 EDT. Figure~\ref{fig:Pixhawk-GPS_validation} shows position measurements from each Pixhawk-GPS module as well as corresponding position errors relative to the surveyed marker. Results show that the total error of Pixhawk-GPS modules is on average 1.87 $\pm$ 0.68~m.
 
\begin{table}[h!] 
\caption{ Error analysis comparing Pixhawk-GPS position measurements and surveyed marker.}
\centering
\resizebox{\textwidth}{!}{%
  \begin{tabular}{ccccccccccccccc}
    \toprule
    \multirow{2}{*}{No.} &\multirow{2}{*}{Samples}  &\multirow{2}{*}{Duration UTC}&
      \multicolumn{3}{c}{ RMSE }\\ 
      \cline{4-6} \
       &&&  East-West & North-South& Total\\
      \toprule \hline
  1 & 48,805 &20:14-22:56 & 0.62 m & 0.68 m & 0.92 m \\ 
  2 &48,485 &20:14-22:56 & 0.66 m& 0.96 m & 1.17 m \\  
  3 & 48,924&20:14-22:56 &1.64 m& 1.32 m & 2.11 m \\
  4 & 48,219&20:14-22:56 & 2.21 m& 1.42 m & 2.62 m\\
  5 &48,997 &20:14-22:56 &1.72 m& 1.63 m & 2.37 m\\
  6 & 48,556&20:14-22:56 &1.84 m& 0.79 m  & 2.0 m\\
  \hline
  mean $\pm$ stdev & 291,986 total  & - & 1.45 $\pm$ 0.66 m& 1.13 $\pm$ 0.38 m & 1.87 $\pm$ 0.68  m\\\midrule
  \bottomrule
  \end{tabular}}
 
  \label{table:results_pixhawk-GPS_validation}
  \end{table}

\subsection{GPS Performance Comparison}
\label{s:GPS_performance_comparison}

Following the validation of Pixhawk-GPS modules next to a surveyed marker, the performance of the Pixhawk-GPS module and SPOT Trace GPS systems tracking drifting assets were compared. This analysis involved quantifying for all manikin releases the RMSE of position measurements obtained from the Pixhawk-GPS and SPOT Trace GPS systems,
\begin{equation}
    \delta r_{S/P} = \sqrt{\frac{1}{N}\sum_{k = 1}^{N}\|\bm{x}_\mathrm{P}(k)-\bm{x}_\mathrm{S}(k)\|^2}
 \label{delta_r_sp}
 \end{equation}
where $\bm{x}_\mathrm{P}$ and $\bm{x}_\mathrm{S}$ are the horizontal position vectors recorded by the Pixhawk and SPOT Tracker GPS systems and $N$ is the number of measurements in the sample. For this assessment, position measurements from the Pixhawk-GPS modules were averaged and interpolated to match the 5-minute resolution of SPOT Trace GPS measurements. Based on East-West and North-South position measurements, shown in Table~\ref{table:results_pixhawk-GPS_SPOT_Trace_GPS}, the RMSE of total displacement exceed 20~m on average. These results suggest that the Pixhawk-GPS modules can track assets with higher resolution and accuracy. 

\begin{table}[tbh!]\small
\centering
  \caption{Error analysis comparing Pixhawk-GPS modules and SPOT Trace GPS systems. }
\resizebox{\textwidth}{!}{%
  \begin{tabular}{ccccccccccccccc}
    \toprule
    \multirow{2}{*}{Track} &\multirow{2}{*}{Samples}  &\multirow{2}{*}{Duration UTC}&\multirow{2}{*}{Distance}&
      \multicolumn{3}{c}{ RMSE }\\ 
      \cline{5-7} \
       &&&&  East-West & North-South& Total\\
      \toprule \hline
  1 & 70 &13:15-19:15 &9.65 km&20.30 m & 8.73 m & 11.85 m \\ 
  2 &45 &13:21-18:51 &10.88 km&8.94 m& 22.36 m & 24.08 m \\  
  3 & 20&14:09-18:35 &6.48 km&12.68 m& 5.78 m & 13.94 m \\
  4 & 25&14:05-16:10 &3.48 km&18.95 m& 24.18 m & 30.72 m\\
  5 &7 &14:41-17:51 &6.87 km&18.58 m& 11.90 m & 22.06 m\\
  6 & 14&14:44-17:04 &7.37 km&21.89 m& 6.10 m  & 22.73 m\\
  \hline
  mean $\pm$ stdev & 181 total  & - & 7.5 $\pm$ 2.6 km & 17 $\pm$ 5.0 m& 13 $\pm$ 8.1 m & 21 $\pm$ 7.0  m\\\hline
  \bottomrule
  \end{tabular}}
  \label{table:results_pixhawk-GPS_SPOT_Trace_GPS}
  \end{table}

 \section{Quadrotor Wind Estimation Framework }
\label{as:quadrotor_wind_estimation_framework}
Wind velocity was measured over drifting objects with a quadrotor employing the model-based wind estimation algorithm presented in~\cite{gonzalez2019sensing}. This approach to wind estimation does not require the quadrotor used for wind estimation to carry a dedicated flow sensor and air data system, which can shorten the flight duration of the quadrotor significantly as a result of the added weight. Instead, wind velocity measurements are inferred from the quadrotor's dynamic response to wind perturbation using a model-based state estimator.  The quadrotor model used for wind estimation is a linear time-invariant model characterizing the quadrotor's rigid-body flight dynamics in hovering flight. The rigid-body model was characterized using system identification experiments presented in~\cite{gonzalez2019sensing,gonzalez2020wind} employing the methodology described in~\cite{klein2006aircraft}. The accuracy of quadrotor wind estimates was then assessed through field experiments next to conventional atmospheric sensors. 

\subsection{Validation of Quadrotor Wind Estimates}
\label{ass:validationf_of_quadrotor_wind_estimates}

\begin{figure}[h!]
\centering
\begin{subfigure}[b]{0.45\textwidth}
         \centering
         \includegraphics[width=5.8cm]{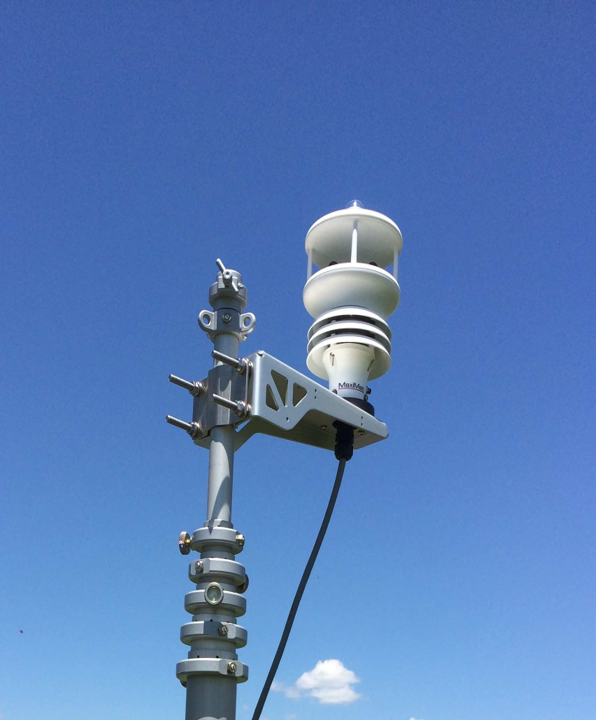}
         \caption{}
         \label{subfig:sonic_anemometer}
\end{subfigure}\hfill
\begin{subfigure}[b]{0.45\textwidth}
         \centering
         \includegraphics[width=\textwidth]{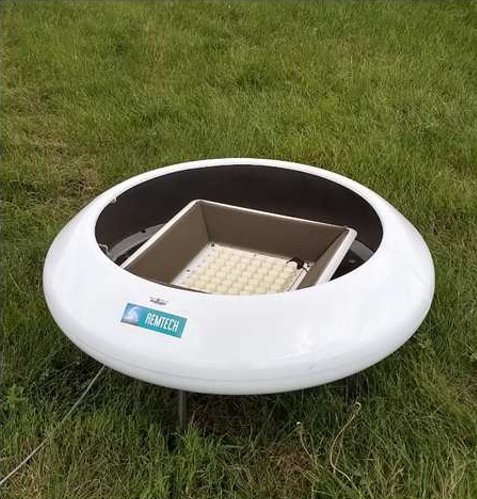}
         \caption{}
         \label{subfig:SoDAR}
\end{subfigure}
 \caption{ Ground-based atmospheric sensors used to validate quadrotor wind estimates. a) WindSonic anemometer mounted on a 10-m telescoping tower. b) The Remtech PA-0 SoDAR capable of profiling wind velocity from 10 to 200~ASL.}
    \label{fig:atmospheric_sensors}
\end{figure}

Flight experiments to validate quadrotor wind estimates were conducted in an open field next to the Virginia Tech Kentland Experimental Aerial Systems (KEAS) Laboratory. The validation procedure involved flying the quadrotor next to the sonic anemometer and SoDAR sensor shown in Figure~\ref{fig:atmospheric_sensors} to measure wind velocity simultaneously. The sonic anemometer used in validation experiments was mounted on a telescoping tower 10 m AGL. The SoDAR sensor, on the other hand, was fixed at the ground. Performance characteristics of the sonic anemometer and SoDAR used in validation experiments are shown in Table~\ref{table:independent_sensors}.
Results from validation experiments were used to determine the accuracy of quadrotor wind estimates using sonic anemometer and SoDar wind observations as ground truth. 

\begin{table}[h!]
\caption{Temporal resolution and accuracy of wind sensors used in validation experiments.}

\centering
  \begin{tabular}{ccccccccccccccc}
    \toprule
    \multirow{2}{*}{Make/Model} &Temporal&
      \multicolumn{2}{c}{ Accuracy }\\
      \cline{3-4} \
       & \small{Resolution} & \small{Wind Speed}  & \small{ Wind Direction}    \\
      \midrule \midrule
\small Remtech PA-0 SoDAR  & \small 300 s& $<$ 0.2 m/s above 6 m/s & $ 3^\circ$ above 2 m/s  \\
\small Gill GMX541 & \small 0.01 s &  $\pm 3.0\%$ to 40 m/s&$ \pm 3^\circ$ to 40 m/s \\\hline
  \bottomrule
  \end{tabular}
  
  \label{table:independent_sensors}
  \end{table}


  \subsection{Wind Estimation Validation Results}
\label{ass:wind_estimation_experimental_validation}

\begin{figure}[h!]
\centering
\begin{subfigure}[b]{0.9\textwidth}
         \centering
         \includegraphics[width=\textwidth]{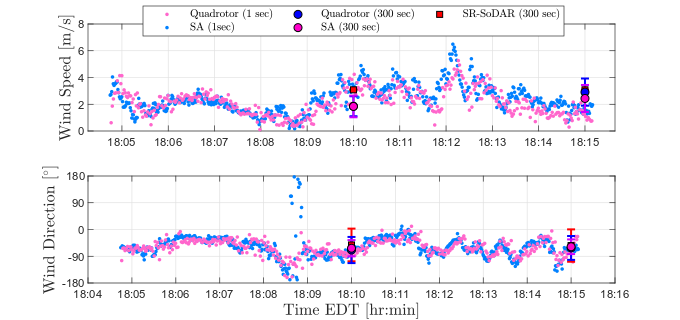}
         \caption{}
         \label{subfig:HoverB}
\end{subfigure}
\begin{subfigure}[b]{0.9\textwidth}
         \centering
         \includegraphics[width=\textwidth]{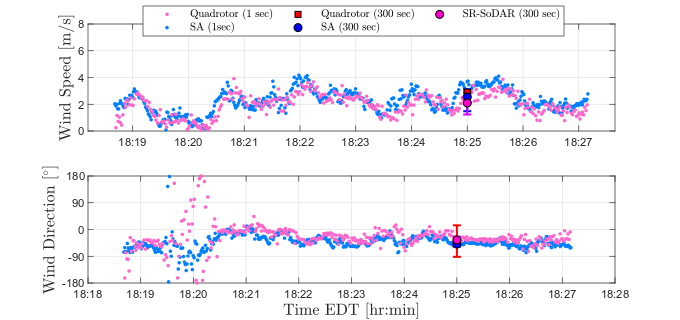}
         \caption{}
         \label{subfig:Hover1818}
\end{subfigure}
\caption{  Wind speed and direction from the sonic anemometer, Remtech SoDAR, and quadrotor at 10~m above ground level (AGL).}
  	\label{fig:hover_measurements}
\end{figure}

 The quadrotor wind estimation algorithm was validated via field experiments next to the sonic anemometer and SoDAR sensor described in Appendix~\ref{ass:validationf_of_quadrotor_wind_estimates}. Validation results for three flights taking place between 18:17 and 20:17  are shown in Figure~\ref{fig:hover_measurements}. Prevailing wind conditions during this time period were from the northwest direction with wind speed varying between 1~m/s and 6~m/s. How well wind measurements from the quadrotor and atmospheric sensors agreed was determined by quantifying both the root mean squared error (RMSE) and mean bias error (MBE). Mean absolute error (MAE) of MBE and RMSE values, shown in Table~\ref{table:SA_SRSoDAR_bias}, demonstrate quadrotor and sonic anemometer (SA) measurements to be within mean absolute errors of 0.3~m/s for wind speed and $9.9^\circ$ for wind directions. Quadrotor and SoDAR comparisons, on the other hand, show wind speed and wind direction measurements to agree within mean absolute errors of 1.0~m/s and $9.9^\circ$, respectively. Hence, quadrotor wind estimates were found to be comparable to wind observations from ground-based atmospheric wind sensors for the range of wind conditions experienced during validation experiments. 
 
Validation results of quadrotor wind estimates suggest that the uncertainty of leeway parameters associated with wind estimation errors to be small when wind speed conditions are below 6~m/s. However, this range of wind conditions is not sufficiently representative of the entire range of wind conditions at sea. For this reason, more validation experiments are required to characterize how the error of multirotor wind estimates increases with intensity of wind speed conditions. Knowing this relationship is critical toward understanding how the uncertainty of leeway parameters will be affected by error in multirotor wind measurements.
 
\begin{table}[h!]
 \caption{ Comparison of five-minute averages of wind speed and wind direction observations collected from the 
 quadrotor, SA, and SoDAR at 10 m AGL on June, 5th 2018 from 18:05 to 20:17~EDT.}
\centering
\resizebox{\textwidth}{!}{%
  \begin{tabular}{cccccccccccccc}
    \toprule
     Flight Time& & \multicolumn{2}{c}{Wind Speed MBE}&& Wind Speed RMSE &&
      \multicolumn{2}{c}{Wind Direction MBE } & &Wind Direction RMSE\\
      \cline{3-4} \cline{6-6}\cline{8-9} \cline{11-11} \\[-1em] EDT &&SA &  SR-SoDAR& &  SA & &  SA&SR-SoDAR&&SA   \\
      \toprule\midrule
  \multirow{2}{*}{18:05-18:15} && 0.0 m/s & 1.2 m/s && \multirow{2}{*}{0.8 m/s} && $7.0^\circ$  &$-4.0^\circ$ &&\multirow{2}{*}{$57.6^\circ$} \\ & & 0.5 m/s&0.7 m/s && &&$4.0^\circ$&$-6.0^\circ$ 
  \\ \midrule
   18:19-18:27 &&-0.5 m/s&-0.8 m/s&& 0.6 m/s&&$4.7^\circ$&$12.7^\circ$&& $260.6^\circ$  \\ \midrule 
   20:08-20:17&& - 0.1 m/s&-1.2 m/s&& 0.4 m/s&&$-23.6^\circ$ &$-17.0^\circ$ && $374.5^\circ$
   \\
        \midrule
       \multicolumn{2}{c}{\textbf{MAE}} & 0.3 m/s &  1.0 m/s && 0.6 m/s&&$9.9^\circ$ & $9.9^\circ$&& 230.9 \\\midrule
    \bottomrule
  \end{tabular}}

    \label{table:SA_SRSoDAR_bias}
  \end{table}

\bibliographystyle{unsrt}
 \bibliography{ref1}
\end{document}